\documentclass[a4paper,aps,pre,superscriptaddress,floatfix,nofootinbib,twocolumn,longbibliography]{revtex4-1}

\usepackage{bm}
\usepackage{bbm}
\usepackage{bbold}
\usepackage{bbm}
\usepackage[pdftex]{graphicx}
\usepackage{latexsym,amsmath,verbatim,amssymb,txfonts}
\usepackage{color}
\usepackage{rotating}
\usepackage{verbatim}
\usepackage{multirow}
\usepackage[english]{babel}
\usepackage{comment}

\begin{document}
\def\be{\begin{equation}}
\def\ee{\end{equation}}

\def\bc{\begin{center}}
\def\ec{\end{center}}
\def\bea{\begin{eqnarray}}
\def\eea{\end{eqnarray}}
\newcommand{\avg}[1]{\langle{#1}\rangle}
\newcommand{\Avg}[1]{\left\langle{#1}\right\rangle}
\newcommand{\mi}{\mathrm{i}} %% roman "i"
\newcommand{\di}{i}    
\def\ie{\textit{i.e.}}
\def\etal{\textit{et al.}}
\def\m{\vec{m}}
\def\G{\mathcal{G}}

\title{Statistical physics of  exchangeable sparse simple networks,\\  multiplex networks and simplicial complexes}

\author{Ginestra Bianconi}
\affiliation{School of Mathematical Sciences, Queen Mary University of London, London, E1 4NS, United Kingdom}
\affiliation{The Alan Turing Institute, The British Library, London NW1 2DB, United Kingdom}
\email{ginestra.bianconi@gmail.com}

\begin{abstract}
Exchangeability is a desired statistical property of network ensembles requiring their invariance  upon relabelling of the nodes. However combining sparsity of  network ensembles with exchangeability is challenging. Here we propose a statistical physics framework and a Metropolis-Hastings algorithm defining exchangeable sparse network ensembles.  The model generates networks with heterogeneous degree distributions by enforcing only {\em global constraints}  while existing (non exchangeable) exponential random graphs enforce an extensive number of {\em local constraints}. This very general theoretical framework to  describe exchangeable networks is here first formulated for uncorrelated simple networks and then it is extended to treat simple networks with degree correlations,  directed networks,  bipartite networks and  generalized network structures including multiplex networks and simplicial complexes. In particular here we   formulate and treat both uncorrelated and correlated exchangeable ensembles of  simplicial complexes using statistical mechanics approaches.
\end{abstract}

\maketitle
\section{Introduction}
Networks constitute the architecture of the vast majority of complex systems ranging from the brain to finance    \cite{Book_Laszlo,Book_Newman}.
Maximum entropy network ensembles \cite{park2004statistical,park2004solution,cimini2019statistical,peixoto2014hierarchical,anand2009entropy,bianconi2013statistical,radicchi2020classical,peixoto2012entropy,orsini2015quantifying,coolen2017generating,kahle2009complexity,pessoa2021entropic,del2010efficient,garlaschelli2008maximum,sagarra2013statistical,krioukov2010hyperbolic} and in general information theory and modeling frameworks \cite{kharel2021degree,de2016spectral,ghavasieh2020statistical,santoro2020algorithmic}    are key to  analyze such realistic networks, and  can be used   for a wide variety of applications.
Due to the deep relation between information theory and statistical mechanics \cite{jaynes1957information}, maximum entropy network ensembles can  to large extent be treated as traditional statistical mechanics ensembles. Indeed recently it has been shown in Ref. \cite{anand2009entropy} that network ensembles can be distinguished between canonical and microcanonical network ensembles enforcing respectively soft and hard constraints. For instance Erd\H{o}s and R\'enyi  networks of $N$ nodes can enforce either a given total number of links $L$ (giving rise to the  $G(N,L)$ ensemble) or a given expected number of links
(giving rise to the ensemble $G(N,p)$ ensemble where $p$ is the probability that any two links are connected).
Erd\H{o}s and R\'enyi random networks are certainly important, however in many applications it is observed that nodes have heterogeneous degree distribution \cite{BA}, typically deviating from the Poisson degree distribution of Erd\H{o}s-R\'enyi networks in the sparse network regime.
In order to treat network ensembles with heterogeneous degree distribution, exponential random graphs \cite{park2004statistical,park2004solution,horvat2015reducing} are considered instead. Exponential random graphs are canonical network ensembles  enforcing a given expected degree sequence. While the Erd\H{o}s R\'enyi  ensembles impose a  single global constraint such as the expected total number of links, exponential random graphs enforce an extensive number of local constraints each given by the expected degree of a single node of the network.
This feature of exponential random graphs makes these ensembles significantly different from the Erd\H{o}s R\'enyi ensembles. The first main difference is  that these ensembles are not any more equivalent to their conjugated microcanonical network ensemble \cite{anand2010gibbs} (the configuration model) which enforces a given degree sequence of the network. The second main difference is that these ensembles are not any more exchangeable.
Exchangeability is a notion originally introduced by de Finetti \cite{deFinetti} whose  theorem states that a sequence of random variables $X_1,X_2\ldots $  is exchangeable if and only if there exists a random probability measure $\Theta$ such
that the $X_i$ are conditionally identically independent variables given $\Theta$. This notion has been then extended to $2$-arrays, (i.e.  networks) for which the Aldous-Hoover theorem applies 
\cite{Aldous,Hoover,lovasz2012large,caron2017sparse}.

Exchangeability is a desired statistical property of network ensembles that ensure  invariance of the model upon relabelling of the nodes. {  In network sampling, when the labels of the nodes depend on the sampling order, exchangeability ensures that the marginal probability of a link is unchanged if nodes are sampled in a different order}. Together with projectivity,  implying that the marginal probability of a link does not change if the network size increases, or if a part of the network is hidden, exchangeability is a  fundamental  property of network models that allows their  reliable use for sampling and for preserving privacy when processing real network data \cite{borgs2015private,borgs2018revealing,kartun2018sparse}.
In the dense network regime, graphons \cite{lovasz2012large,wolfe2013nonparametric} have been shown to be exchangeable and projective and are known  to allow a well defined infinite graph limit \cite{Aldous,Hoover}. Graphons are dense in the sense that they have a number of links of the same order of the number of nodes to the power two, i.e. $L={O}(N^2)$. However this regime is seldom encountered in real networks.  The  mathematics literature has recently proposed several approaches to face the challenge of modelling exchangeable networks with a number of links that scales like $L=O(N^{1+\hat{\alpha}})$ with $0<\hat{\alpha}<1$ \cite{caron2017sparse,veitch2015class,borgs2016sparse,veitch2019sampling} and to define ways to define the infinite network limit \cite{borgs2016sparse} for such models. All these approaches are based on  point processes on  $\mathbb{R}_+^2$.

In this paper we propose a statistical physics approach to model sparse exchangeable network ensembles  of a given number of nodes $N$ and a number of links that scales as $L=O(N)$. Therefore these network ensembles cover the  scaling regime $L=O(N)$ which is  important for the vast majority of applications. The exchangeable network ensembles are  Hamiltonian and are not based on a point processes. These ensembles generate networks with given heterogeneous degree distribution $p(k)$ imposing only two global constraints: the energy (expressing the value of a global exchangeable Hamiltonian of the network ensemble) and the total number of links. 
The proposed exchangeable sparse network ensembles  have the property that each link of the network has the same marginal probability, still the network display and heterogeneous degree distribution. Moreover  the probability that two nodes are connected, when conditioned on their degrees  reduces to the probability of the exponential random graph in the uncorrelated limit. 
{ 
Note that the model is not projective and in particular  the marginal distribution of a link is actually dependent on the network size. Therefore we do not consider the network generated in the limit  $N\to\infty$, instead we take an equilibrium statistical mechanics approach and we consider $N$ finite but large.
Indeed our result do not contradict the Aldous-Hoover theorem \cite{Aldous,Hoover,lovasz2012large,caron2017sparse} that   states that infinitely exchangeable sparse
networks are  empty. In fact in for our model in the limit $N\to \infty$ limit the  marginal probability of any link goes to zero.
Although this can be considered a shortcoming with respect to graphons, this does not limit the applicability of the model. Indeed many widely used network models are not projective including the Erd\H{o}s R\'eny networks and the exponential random graphs.
}
The model is here  simulated with a constructive Metropolis algorithm and is extended to network models with degree correlations, to directed, bipartite networks  and to generalized network structures such as multiplex networks      \cite{bianconi2013statistical,bianconi2018multilayer,boccaletti2014structure} and simplicial complexes \cite{bianconi2021higher,battiston2021physics,courtney2016generalized,ghoshal2009random,battiston2020networks,wegner2021atomic}.
\section{Fundamental properties of exchangeable network ensembles}
A network ensemble is exchangeable if the   probability
$\mathbb{P}(G)$ of a network $G = (V, E)$ is independent on the nodes labels, i.e.,
\bea
\mathbb{P}(G) = \mathbb{P}(\tilde{G}), 
\eea
{  for any labelled network $\tilde{G}$  obtained from the network $G$ by permuting the nodes labels, in particular this includes all labelled networks $\tilde{G}$ isomorphic to $G$. }\\
{  Assuming that each labelled network $G$ is uniquely determined by the adjacency matrix ${\bf a}$, and that the ensemble is determined by the probability $P(G)=P({\bf a})$, we define the marginal probability of a link $(i,j)$ as 
\bea
p_{ij}=\sum_{{\bf a}}a_{ij}P({\bf a}).
\eea}
From the definition of exchangeable network ensemble it  follows that in an exchangeable network ensemble the  marginal probability $p_{ij}$ of a link between node $i$ and node $j$  must  be invariant under any permutation $\sigma$  of the node labels, i.e.
\bea
p_{ij} = p_{\sigma(i),\sigma(j)}.
\eea
{  Note however that this is a necessary condition for exchangeable network ensembles  and not a sufficient condition. Indeed  knowing the marginal probabilities of the link of a network might not be enough to determine general network ensembles  for which  $P(G)$ that does not factorize into independent link probabilities,  (see for instance  the $2$-star or the Strauss model \cite{Book_Newman}).}
In an exponential random graph ensemble with given expected degree sequence ${\bf k}=\{k_1,k_2,\ldots, k_N\}$ with $k_i\leq K\ll K_S=\sqrt{\avg{k}N}$  the marginal distribution $p_{ij}$ takes the well celebrated expression
\bea
p_{ij}=\frac{k_ik_j}{\avg{k}N},
\eea
where $\avg{k}N=\sum_{i=1}^Nk_i=2L$ is twice the  expected total  number of links of the network.
This  network ensemble  is not exchangeable, unless the expected degree of each node is the same. Indeed the marginal probability $p_{ij}$ is not invariant upon permutation of the node labels if the expected degree distribution is heterogeneous. Only in the case in which the expected degree of each node is the same $k_i=\avg{k}$ we recover the exchangeable expression of the marginal probability of a sparse Poisson Erd\H{o}s and R\'enyi network $p_{ij}=\avg{k}/N$. Note that both the Erd\H{o}s and R\'enyi network and the exponential random graph are not projective. This is an immediate consequence of the fact that the marginal probability $p_{ij}$ of the link $(i,j)$ depends on the network size $N$. Therefore upon addition of new nodes, leading to an increase of $N$, the marginal probability between two previously existent nodes changes.

In the following section we will propose an exchangeable sparse network ensemble that imposes a heterogeneous expected degree distribution $p(k)$ enforcing only two global constraints: the energy of the ensemble and the total number of links.
In this ensemble the marginal probability of a link between node $i$ and node $j$ is given by the exchangeable expression, 
\bea
p_{ij}=\sum_{k,k'}p(k)p(k')\frac{kk'}{\avg{k}N}=\frac{\avg{k}}{N}.
\label{marginal}
\eea
In other words the marginal probability of any link is the same for every link, but when it is conditioned to the degree of the two linked nodes is given by the uncorrelated network expression
\bea
p_{ij|k_i=k,k_j=k'}=p(k,k')=\frac{kk'}{\avg{k}N}.
\label{conditioned}
\eea
Therefore this ensemble has the same marginal of the Erd\H{o}s and R\'enyi network but it can generate uncorrelated networks with heterogeneous degree distribution $p(k)$.\\
{  Note that as $N\to \infty$ the marginal probability given by Eq. (\ref{marginal}) vanishes. This implies that  our exchangeable sparse network ensemble does not contradict  Aldous-Hoover theorem according to which any sparse infinite exchangeable network should vanish. However our approach  provides finite exchangeable network models with given degree distribution for any large but finite value of $N$.} \\
As we will see in subsequent sections this network ensemble can be extended to sparse simple networks with degree correlations, to directed, bipartite networks and to generalized network structures.
In all these cases the marginal probability of an interaction is the same for any possible interaction of the network, yet the exchangeable ensembles can give rise to networks with very heterogeneous topology.

\section{Exchangeable sparse simple network ensembles}
In this section our goal is to construct ensembes of sparse simple  network  of $N$ nodes with degree distribution $p(k)$. Here by {\em sparse} we imply that these networks  display a minimum degree $\hat{m}$ and  a maximum degree (cutoff) $K$ much smaller than the structural cutoff $K_S$, i.e. $k_i\leq  K\ll K_S=\sqrt{\avg{k}N}$ for all $i\in \{1,2,3,\ldots, N\}$ with $\avg{k}$ indicating the expected value over the $p(k)$ distribution, $\avg{k}=\sum_k kp(k)$. We assume that the nodes of the network can change their degree and  we assign to each possible degree sequence ${\bf k}=\{k_1,k_2,\ldots, k_N\}$ of the network the probability
\bea
P({\bf k})=\prod_{i=1}^N\left[p(k_i)\theta(K-k_i)\theta(k_i-\hat{m})\right].
\eea
where $\theta(x)$ indicates the Heaviside function with  $\theta(x)=1$ for $x\geq 0$ and $\theta(x)=0$ otherwise.
Therefore the probability of a degree sequence results from the product of the probability that each node has the observed degree.
For keeping the model general we assume that the minimum degree of the network must be equal or greater than $\hat{m}$. For instance if we want to impose a power-law degree distribution $p(k)=ck^{-\gamma}$ this allow us to impose a minimum degree $\hat{m}\geq 1$ and to exclude isolated nodes for which $p(k)$ is not defined. However also $\hat{m}=0$ is allowed as long as $p(k)$ is well defined for $0\leq k\leq K.$
In order to build an exchangeable network ensemble we need to define a probability $\mathbb{P}(G)$ for any possible network $G=(V,E)$ in the ensemble described by the adjacency matrix ${\bf a}$. In order to ensure sparsity we impose that the total number of links $L$ is fixed and given by  $L=\avg{k}N/2$  and we impose that the probability of getting a degree sequence ${\bf k}$ is $P({\bf k})$. 
Since the number of networks $\mathcal{N}$ with given degree sequence ${\bf k}$ can be expressed in terms of the entropy $\Sigma({\bf k})$ of networks with degree sequence ${\bf k}$ as \cite{anand2009entropy} $\mathcal{N}=\exp\left[{\Sigma({\bf k})}\right]$, 
the probability of each single network $G$ displaying a degree sequence ${\bf k}$  is therefore taken to be
\bea
\mathbb{P}(G)=P({\bf k})e^{-\Sigma({\bf k})}\delta\left(L,\sum_{i<j}a_{ij}\right),
\label{P}
\eea
where $\delta(x,y)$ indicates the Kronecker delta.
For sparse networks  the entropy $\Sigma({\bf k})$ of networks with given degree sequence with $k_i\ll K_S$ obeys the Bender-Canfield formula \cite{bender,bianconi2008entropies,anand2009entropy,anand2010gibbs}
\bea
{\Sigma}({\bf k})= \ln \left(\frac{(2L)!!}{\prod_{i=1}^Nk_i!}\right)+o(N)
\label{S}
\eea
 where in Eq. (\ref{P}) and (\ref{S}) we indicate with ${\bf k}=\{k_1,k_2,\ldots, k_N\}$ the degree sequence with    $k_i$, the degree of node $i$, given by 
$k_i=\sum_{j=1}^Na_{ij}.$
Note that the sparse exchangeable network ensemble greatly differs from the network ensemble with given expected degree sequence because in the exchangeable ensemble the constraints are global and not local.
Indeed the expression for $\mathbb{P}(G)$ can be also given by 
\bea
\hspace*{-7mm}\mathbb{P}(G)=e^{-H(G)}\delta\left(L,\sum_{i<j}a_{ij}\right)\theta\left(K-\max_{i=1,\ldots, N}k_i\right)\theta\left(\min_{i=1,\ldots, N}k_i-\hat{m}\right),
\eea
with {\em Hamiltonian} $H(G)$ given by 
\bea
H(G)=-\sum_{i=1}^N\ln p(k_i)+\Sigma({\bf k}).
\eea
Using Eq. (\ref{S}) for $\Sigma({\bf k})$ we can derive the explicit expression for $H(G)$:
\bea
H(G)=-\sum_{i=1}^N\ln [p(k_i)k_i!]+\ln\left((2L)!!\right)
\label{H}
\eea
The Hamiltonian $H(G)$ is clearly a global variable that depends on all the nodes of the network where each node is treated on equal footing. Therefore the expression of the Hamiltonian is clearly exchangeable as it is invariant upon a permutation of the node labels. 
In order to show that  the marginal distribution is given by Eq. (\ref{marginal}) we solve the model using saddle point method applied to a free-energy expressed in terms of a functional order parameter.
In order to perform this calculation, let us write the probability $\mathbb{P}(G)$ as
\bea
\hspace{-7mm}\mathbb{P}(G)=\frac{1}{(2L)!!}\sum^{\prime}_{\bf k}\prod_{i=1}^N\left[k_i!p(k_i)\delta\left(k_i,\sum_{j=1}^Na_{ij}\right)\right]\delta\left(\sum_{i,j}a_{ij},L\right),
\label{14}
\eea
where $\sum_{{\bf k}}^{\prime}$ indicates the sum over all possible degree sequences with a maximum degree equal or smaller than $K$ and a minimum degree greater or equal than $\hat{m}$.
Expressing the Kronecker deltas in Eq. (\ref{14}) with their integral form 
\bea
\delta(x,y)=\frac{1}{2\pi}\int_{-\pi}^{\pi}d\omega e^{\mi\omega(x-y)},
\label{delta}
\eea
the partition function $Z=Z(h)$ can  be expressed as
\bea
Z&=&\sum_{{\bf a}}\mathbb{P}(G)e^{-h\sum_{i<j}a_{ij}}\nonumber \\
&=&\frac{1}{(2L)!!}\sum_{{\bf a}}\sum_{\bf k}^{\prime}\int {{\mathcal D}\omega}\int \frac{d\lambda}{2\pi} e^{G(\lambda,\bm{\omega},{\bf k},h)},
\label{Zsimple}
\eea
with 
\bea
G(\lambda,\bm{\omega},{\bf k},h)&=&\sum_{i=1}^N[\mi\omega_i k_i+\ln (k_i!p(k_i))]+\mi\lambda L\nonumber \\
&&+\frac{1}{2}\sum_{i,j}\ln (1+e^{-\mi\lambda-\mi\omega_i-\mi\omega_j-h}),
\eea
and ${\mathcal D}\omega=\prod_{i=1}^N [d\omega_i/(2\pi)]$.
Let us now  introduce the functional order parameter indicating the density of nodes with degree $k_i=k$ and with $\omega_i=\omega$, \cite{courtney2016generalized,bianconi2008entropies,monasson1997statistical}
\bea
c(\omega,k)=\frac{1}{N}\sum_{i=1}^N\delta(\omega-\omega_i)\delta(k,k_i).
\eea
By calculating the partition function $Z$ in the sparse regime (i.e. $K\ll K_S$) with the saddle point method (see Appendix A) we can derive for  the functional order parameter $c(\omega, k)$ when $h\to 0$ the expression 
\bea
c(\omega,k)=\frac{1}{2\pi}p(k)k! e^{i\omega k+e^{-i\omega}}.
\label{c}
\eea
This implies that  the density of nodes of degree $k$ is given by 
\bea
\int d\omega c(\omega,k)=p(k).
\eea
Therefore  the degree of each node is fluctuating, but in the large network limit the degree distribution is given by $p(k)$ as desired.
We are now in the position to evaluate the marginal probability $p_{ij}$ of a link between node $i$ and node $j$. A straightforward calculation (see Appendix A) leads to the expression of the marginal probability $p_{ij}=\avg{a_{ij}}$ in terms of the functional order parameter $c(\omega,k)$ leading to
\bea
p_{ij}=\frac{1}{\avg{k}N}\sum_{\hat{m}\leq k\leq K,\hat{m}\leq k'\leq K}\int_{-\pi}^{\pi} \frac{d\omega}{2\pi} \int_{-\pi}^{\pi} \frac{d\omega^{\prime}}{2\pi}c(\omega,k)c(\omega',k')e^{-\mi\omega-\mi\omega'}.\nonumber
\eea
From this expression, using Eq. (\ref{c}) it follows immediately the expression for the marginal given by Eq. (\ref{marginal}) leading to $p_{ij}=\avg{k}/N$ also if the marginal probability conditioned on the nodes degree (see Appendix A for a detailed derivation) is given by Eq. (\ref{conditioned}).
Therefore the marginal probability $p_{ij}=\avg{k}/N$ is the same for every node of the network and it is equal to the marginal probability in a  Poisson Erd\H{o}s R\'enyi network, but the degree distribution is $p(k)$, i.e. it can significantly differ from a Poisson distribution.
\section{Metropolis-Hastings algorithm} 
\begin{figure}[htbp!]
		\centering
	\includegraphics[width=0.95\columnwidth]{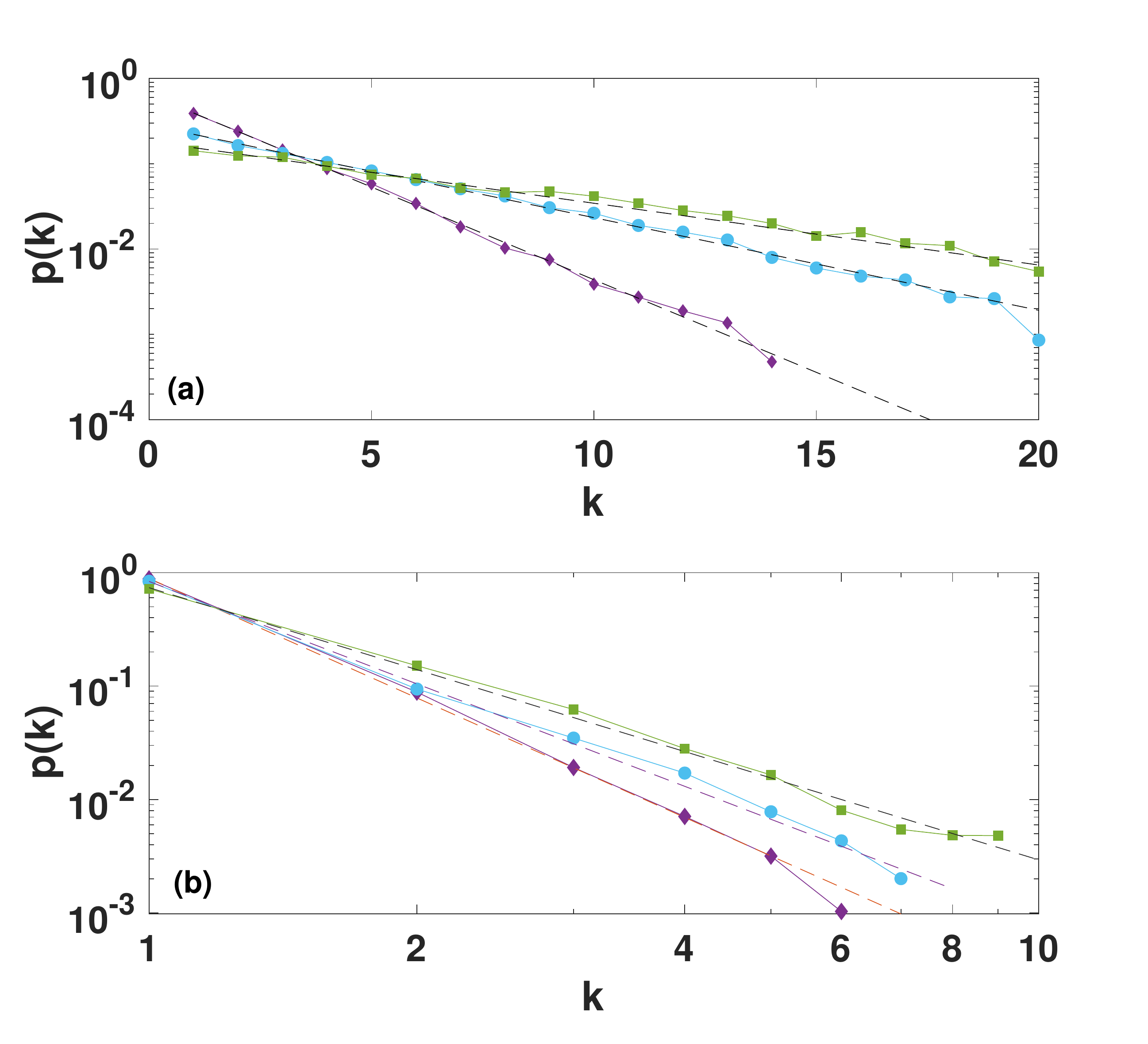}
	\caption{Degree distributions of the exchangeable uncorrelated network ensembles generated by the Metropolis-Hastings algorithm. Panel (a) shows the case of exponential degree distributions $p(k)=ce^{-k/k_0}$ with $k_0=2$, (green squares) $k_0=4$ (cyan circles) and $k_0=6$ (purple diamonds), panel (b) shows the case of power-law degree distributions with $p(k)=ck^{-\gamma}$ and $\gamma=2.4$ (green squares) $\gamma=3.0$ (cyan circles) and $\gamma=3.5$ (purple diamonds). The dashed lines indicate the theoretical expectation. In all simulations the networks have $N=2000$ nodes. }
	\label{fig1}
	\end{figure}

The exchangeable ensemble of sparse networks can be obtained by  implementing a simple Metropolis-Hastings algorithm using the network Hamiltonian given by Eq.(\ref{H}). The Metropolis-Hastings algorithm for the exchangeable sparse networks is outlined below.
\begin{itemize}
\item[(1)] Start with a network of $N$ nodes having exactly $L=\avg{k}N/2$ links and in which the minimum degree is greater of equal to $\hat{m}$ and the maximum degree is smaller or equal to $K$.
\item[(2)]Iterate the following steps until equilibration:
\begin{itemize}
\item[(i)] Choose randomly a random link $\ell=(i,j)$ between node $i$ and $j$ and choose a pair of random nodes $(i',j')$ not connected by a link. By indicating with ${\bf a}$ the {(symmetric)} adjacency matrix of the network we have $a_{ij}=1$ and $a_{i'j'}=0$. 
\item[(ii)] Let ${\bf a'}$ be the adjacency matrix of the network in which the link $(i, j)$ is removed and the link  $(i',
j')$ is inserted instead. Draw a random number $r$ from a uniform
distribution in $[0, 1]$, i.e. $r\sim U(0, 1)$. 
If $r <\mbox{max}(1,e^{-\Delta H})$
where $\Delta H= H({\bf a})-H({\bf a'})$ and if the move does not violate the conditions
on the minimum and maximum degree of the network, replace ${\bf a}$ by ${\bf a'}$.
\end{itemize}
\end{itemize}
This algorithm can be used to generate exchangeable network ensembles with different degree distributions such as exponential distribution or power-law degree distribution (see Figure $\ref{fig1}$).\\
This approach can be directly extended to treat sparse networks with given degree correlations, directed networks, bipartite networks and also generalized network structures such as multiplex networks and simplicial complexes as we will describe  in the following sections.
\section{Exchangeable network ensembles with degree correlations} 
Degree correlations are important characteristic of networks that have attracted large interest   \cite{pastor2001dynamical,vazquez2002large,maslov2002specificity,srivastava2012correlations}
 because they encode additional network information not captured by the degree distribution.
In particular in the literature different works have been proposed to model {\em labelled} network with given degree correlations  \cite{park2003origin,bianconi2008entropies,coolen2017generating,bassler2015exact}.\\
Here our aim is to  construct  a sparse exchangeable network ensemble of $N$ nodes in which each node has degree $k$ with probability $p(k)$ and every link between a node of degree $k$ and a node of degree $k'$ contributes to the probability of the network $G$ by a factor $Q(k,k')$ where $Q(k,k')=Q(k,k')$ with $Q(k,k')$ independent of $N$. As we will see in the following, by changing the kernel function $Q(k,k')$ it is possible to select the nature of the degree correlations modulating the probability $p(k,k')$ of  observing a link between a node of degree $k$ and a node of degree $k'$. In order to define the exchangeable ensemble of correlated networks, we follow a derivation similar to the one considered in the previous {  section} and we assign to each network $G$ a probability $\mathbb{P}(G)$ given by
\bea
\mathbb{P}(G)&=&\prod_{i<j}[Q(k_i,k_j)]^{a_{ij}}\prod_{i=1}^Np(k_i)e^{-\Sigma({\bf k})}\delta\left(L,\sum_{i<j}a_{ij}\right)\nonumber \\
&&\times \theta\left(K-\max_{i=1,\ldots, N}k_i\right)\theta\left(\hat{m}-\min_{i=1,\ldots, N} k_i\right),
\eea
where $\Sigma({\bf k})$ is the entropy of the ensemble of correlated networks with degree sequence ${\bf k}$. The entropy $\Sigma({\bf k})$ {  has been calculated in Refs.} \cite{bianconi2008entropies,annibale2009tailored} and can be expressed as  
\bea
\Sigma({\bf k})=\ln \left((2L)!! \prod_{i=1}^N\frac{[\gamma(k_i)]^{k_i}}{k_i!}\right)+o(N)
\label{Sc+}
\eea
where $\gamma(k)$ is a function that satisfies 
\bea
\gamma(k)=\frac{1}{\avg{k}}\sum_{k'}Q(k,k')p(k')\frac{k'}{\gamma(k')}.
\label{gamma}
\eea
The probability $\mathbb{P}(G)$ can be also written in Hamiltonian form as 
\bea
\hspace*{-6mm}\mathbb{P}(G)=e^{-H(G)}\delta\left(L,\sum_{i<j}a_{ij}\right)\theta\left(K-\max_{i=1,\ldots, N}k_i\right)\theta\left(\hat{m}-\min_{i=1,\ldots,N} k_i\right)\nonumber 
\eea
where the (exchangeable) Hamiltonian $H(G)$ is given by 
\bea
H(G)=-\sum_{i<j}a_{ij}\ln Q(k_i,k_j)-\sum_{i=1}^N\ln p(k_i)+\Sigma({\bf k}),
\eea
and $\Sigma({\bf k})$ is given by Eq. (\ref{Sc+}).
By studying this correlated network ensemble with  statistical mechanics methods similar to the ones we have used in the case of the exchangeable uncorrrelated network ensemble investigated earlier, we can show (see Appendix B for details) that the degree distribution is $p(k)$ and that the marginal probability of each link can be written as 
\bea
p_{ij}=\sum_{\hat{m}\leq k\leq K,\hat{m}\leq k'\leq K}p(k)p(k')p(k,k')=\frac{\avg{k}}{N},
\label{c1}
\eea
with $p(k,k')$ indicating the probability of a link between a node of degree $k$ and a node of degree $k'$, with
\bea
p_{ij|k_i=k,k_j=k'}=p(k,k')= \frac{1}{\avg{k}N}{Q(k,k')}\frac{kk'}{\gamma(k)\gamma(k')}.
\label{c2}
\eea
Note that for  $Q(k,k')=1$ it follows immediately from Eq. (\ref{gamma}) that $\gamma(k)=1$ and we recover the uncorrelated network ensemble.
Eq. (\ref{c1}) and Eq. (\ref{c2}) clearly reveal the exchangeable nature of this ensemble as the marginal probability of a link is independent of the node labels. However the ensemble retains the ability to model sparse networks with arbitrary degree distribution and degree correlations.

\section{Exchangeable sparse directed network ensembles}
It is well known that a number of real networks including prey-predator interactions in ecology, financial contracts between banks, the World-Wide-Web or  directed online social networks like Twitter, are actually directed. In directed networks  a link $(i,j)$ indicating a directed interaction from node $i$ to node $j$ is distinguished from the link $(j,i)$. For instance the existence of a prey-predator interaction between a species $i$ (prey) and a species $j$ (predator) is not typically reciprocated.
The main difference between simple and directed networks is that in  directed networks the adjacency matrix is not symmetric and for each node we can distinguish between in in-degree and the out-degree. Assuming that $a_{ij}=1$ indicates the presence of a directed link from node $i$ to node $j$, the in-degree and the out-degree of node $i$ can be expressed as 
\bea
k_i^{in}&=&\sum_{j=1}^Na_{ji},\nonumber \\
k_i^{out}&=&\sum_{j=1}^Na_{ij}.
\eea
Several previous work have modelled {\em labelled} directed network ensembles using statistical mechanics approaches  \cite{park2004statistical,bianconi2009entropy,kim2012constructing,roberts2012unbiased,garlaschelli2004patterns}.
 Here we define the exchangeable uncorrelated ensemble of directed networks with joint degree distribution $p_d(k^{in},k^{out})$ indicating the probability that a generic  node $i$ has degree $k_i^{in}=k^{in}$ and $k_i^{out}=k^{out}$. This distribution is arbitrary, but needs to satisfy $\Avg{k^{in}}=\Avg{k^{out}}$.
In order to guarantee sparsity, we assume that the in-degree and the out-degree have a maximum value equal or smaller than $K$ with $K\ll K_s=\sqrt{\Avg{k^{in}}N}$ and that they a minimum value equal or larger than $\hat{m}$.
The exchangeable uncorrelated ensemble of directed networks assigns to each directed network $G$ the probability $\mathbb{P}(G)$ given by 
\bea
\mathbb{P}(G)&=&\prod_{i=1}^Np(k_i^{in},k_i^{out})e^{-\Sigma({\bf k}^{in},{\bf k}^{out})}\delta\left(L,\sum_{i,j}a_{ij}\right)\theta\left(K-\max_{i=1,\ldots, N}k_i^{in}\right)\nonumber \\
&&\times
\theta\left(K-\max_{i=1,\ldots, N}k_i^{out}\right)
\theta\left(\min_{i=1,\ldots, N}k_i^{in}-\hat{m}\right)
\theta\left(\min_{i=1,\ldots, N}k_i^{out}-\hat{m}\right),\nonumber
\eea
where the  entropy $\Sigma({\bf k}^{in},{\bf k}^{out})$  is given by 
\bea
\Sigma({\bf k})=\ln \left[\frac{L!}{\prod_{i=1}^N [k_i^{in}! k_i^{out}!]}\right]+o(N).
\eea
The probability $\mathbb{P}(G)$ admits an Hamiltonian expression as 
\bea
\mathbb{P}(G)&=&e^{-H(G)}\delta\left(L,\sum_{i,j}a_{ij}\right)\theta\left(K-\max_{i=1,\ldots, N}k_i^{in}\right)
\theta\left(K-\max_{i=1,\ldots, N}k_i^{out}\right)\nonumber \\
&&\times
\theta\left(\min_{i=1,\ldots, N}k_i^{in}-\hat{m}\right)
\theta\left(\min_{i=1,\ldots, N}k_i^{out}-\hat{m}\right),
\eea
where the Hamiltonian $H(G)$ of this ensemble is given by
\bea
H(G)=-\sum_{i=1}^N\ln \left(p_d(k_i^{in},k^{out}_i)k_i^{in}!k_i^{out}!\right)+\ln(L!).
\label{directed}
\eea
%\end{document}
The statistical mechanics treatment of this model (see Appendix C) shows that the density of nodes with in-degree $k^{in}$ and out-degree $k^{out}$ is given by the desired joint distribution $p_d(k^{in},k^{out})$ although the marginal probability of each node is equal for each node and given by 
\bea
p_{ij}=\sum_{k^{in},k^{out}}p_{in}(k^{in})p_{out}(k^{out})\frac{k^{in}k^{out}}{\Avg{k^{in}}N},
\label{marginal_directed}
\eea
where 
\bea
p_{in}(k^{in})=\sum_{k^{out}}p_d(k^{in},k^{out}),\nonumber \\
p_{out}(k^{out})=\sum_{k^{in}}p_d(k^{in},k^{out}).
\eea
Note that although the marginal probability is the same for each node the marginal probability of a directed link conditioned on the in-degree and the out-degree of its two end nodes is not, i.e.
\bea
p_{ij|k_i^{out}=k^{out},k_j^{in}=k^{in}}=p(k^{in},k^{out})=\frac{k^{in}k^{out}}{\Avg{k^{in}}N}.
\label{directed_conditioned}
\eea
\section{Exchangeable bipartite network ensembles}
Bipartite networks are another notable class of networks formed by two set of nodes and interactions only exiting between nodes of one class and nodes of  the other class. Example of bipartite networks are mutualistic networks in ecology \cite{bascompte2007plant}, social networks between individuals and taste/opinions, and in general can be used to partition a given set of nodes in different groups \cite{newman2002random}.
Interestingly bipartite networks are also called  factor graphs and are widely used as the architecture supporting  graphical models \cite{mezard2002analytic,monasson1997statistical}.

In this section we consider exchangeable ensembles of bipartite networks formed by two set of nodes $V$ and $U$ with $|V|=N$ and $|U|=M$ with the condition
\bea
M=\alpha N,
\eea
with  $\alpha>0$ being a constant independent of $N$.
We indicate with $i$ the nodes belonging to the set $V$ and with $\mu$ the nodes belonging to the set $U$. The structure of the bipartite network is determined by the $N\times M$ incidence matrix $b_{i\mu}=1$ if there is a link between node $i$ and node $\mu$, otherwise $b_{i\mu}=0$. The degree of the nodes in $V$ and in $U$ is determined from the incidence matrix ${\bf b}$ according to the following equations
\bea
k_i&=&\sum_{\mu=1}^M b_{i\mu},\nonumber \\
q_{\mu}&=&\sum_{i=1}^N b_{i\mu}.
\eea
Here we formulate the exchangeable sparse bipartite network ensemble  designed in order to obtain bipartite networks  in which the nodes  in $V$ have degree distribution $p(k)$ and the nodes in $U$ have degree distribution $\hat{p}(q).$ These distributions can be arbitrary but must obey $N\avg{k}=M\avg{q}$ which implies $\avg{k}=\alpha\avg{q}$. Moreover these ensembles have fixed number of links $L=\avg{k}N$ and the degree $k$ ($q$) of the nodes in $V$ ($U$) has maximum smaller or equal to $K\ll K_s=\sqrt{\Avg{k}N}$ ($\hat{K}\ll K_S=\sqrt{\Avg{k}N}$) and minimum degree greater or smaller than $\tilde{m}$ ($\hat{m}$).
The probability $\mathbb{P}(G)$ for each bipartite network $G$ is taken to be  
\bea
\mathbb{P}(G)&=&\prod_{i=1}^Np(k_i) \prod_{\mu=1}^M\hat{p}(q_{\mu})e^{-\Sigma({\bf k},{\bf q})}\delta\left(L,\sum_{i,\mu}b_{i\mu}\right)\theta\left(K-\max_{i=1,\ldots,N}k_i\right)\nonumber \\
&&
\theta\left(\hat{K}-\max_{\mu=1,\ldots,M} q_{\mu}\right)\theta\left(\min_{i=1,\ldots, N}k_i-\tilde{m}\right)\theta\left(\min_{\mu=1,\ldots, M} q_{\mu}-\hat{m}\right).\nonumber
\eea
The entropy of this ensemble is given by 
\bea
\Sigma({\bf k},{\bf q})=\ln \left[\frac{L!}{\prod_{i=1}^N k_i!\prod_{\mu=1}^M q_\mu!}\right]+o(N)
\eea
This ensemble is Hamiltonian as $\mathbb{P}(G)$ can be expressed as 
\bea
\mathbb{P}(G)&=&e^{-H(G)}\delta\left(L,\sum_{i,\mu}b_{i\mu}\right)\theta\left(K-\max_{i=1,\ldots,N}k_i\right)\nonumber \\&&
\theta\left(\hat{K}-\max_{\mu=1,\ldots,M} q_{\mu}\right)\theta\left(\min_{i=1,\ldots, N}k_i-\tilde{m}\right)\theta\left(\min_{\mu=1,\ldots, M} q_{\mu}-\hat{m}\right)\nonumber
\eea
with the Hamiltonian $H(G)$ given by 
\bea
H(G)=-\sum_{i=1}^N\ln \left(p(k_i)k_i!\right)-\sum_{\mu=1}^M\ln \left(\hat{p}(q_{\mu})q_{\mu}!\right)+\ln( L!)\nonumber\\
\eea
This ensemble produces a network model that can be treated using statistical mechanics methods (see Appendix D) which clearly show that the nodes in $V$ have degree distribution $p(k)$ and the nodes in $U$ have degree distribution $\hat{p}(q)$. Both $p(k)$ and $\hat{p}(q)$ can be heterogeneous even if the marginal of every link is the same for every linkand given by 
\bea
p_{i\mu}=\sum_{k,q}p(k)\hat{p}(q)\frac{kq}{\avg{k}N}.
\label{marginal_bipartite}
\eea
Note that  the marginal probability $p_{i\mu|k_i=k,q_{\mu}=q}=p(k,q)$ of a link ($i,\mu$) conditioned on the degrees $k_i=k$ and $q_{\mu}=q$ in this ensemble is given by
\bea
p_{i\mu|k_i=k,q_{\mu}=q}=p(k,q)=\frac{kq}{\avg{k}N}.
\label{bipartite_conditioned}
\eea
\section{Exchangeable sparse multiplex networks} 
A large variety of complex systems including biological, social networks and infrastructures are better described by multiplex networks  \cite{bianconi2018multilayer,boccaletti2014structure,kiani2021networks} which are formed by a set of nodes connected by two or more networks indicating interactions of different nature and connotations.
The different networks forming the multiplex networks are also called the {\em layers} of the multiplex network.
Different works have investigated ensembles of {\em labelled} multiplex networks with different types of correlations between the layers, with weights of the links and with non-trivial spatial embedding \cite{bianconi2013statistical,menichetti2014weighted,halu2014emergence}.

Here our goal is to propose and investigate the properties of exchangeable sparse multiplex networks.
To this end we can  consider a multiplex network $\vec{G}=(G_1,G_2,\ldots, G_M)$ formed by $M$ layers $\alpha\in \{1,2,\ldots, M\}$ each determined by a adjacency matrix ${\bf a}^{[\alpha]}$ \cite{bianconi2018multilayer}. To keep the discussion simple we will assume that each adjacency matrix is undirected and unweighted. The degree $k_i^{[\alpha]}$ of each node $i$  in layer $\alpha\in\{1,2,\ldots, M\}$ is determined by the equation 
\bea
k_i^{[\alpha]}=\sum_{j=1}^Na_{ij}^{[\alpha]}.
\eea
An important feature of multiplex networks are multilinks $\vec{m}=\left(m^{[1]},m^{[2]},\ldots, m^{[M]}\right)$ (with $m^{[\alpha]}\in \{0,1\}$) \cite{bianconi2013statistical} indicating the pattern of connection between any two nodes. For instance in a duplex network ($M=2$) with two layers indicating mobile phone and email interaction a two nodes are connected by a multilink $(1,0)$ if they only communicate with mobile phone, they are connected by a multilink $(0,1)$ if they only communicate via email and they are connected by a multilink $(1,1)$ if they communicate both via mobile phone and email.
In order to indicate if two nodes $i$ and $j$ are connected by a multilink $\vec{m}$ we can use the {\em multi-adjacency matrices} $A^{\vec{m}}$ \cite{bianconi2013statistical,bianconi2018multilayer} whose element $A_{ij}^{\vec{m}}$ indicates whether node $i$ and node $j$ are connected by a multilink of type $\vec{m}$ ($A_{ij}^{\vec{m}}=1$) or not ($A_{ij}^{\vec{m}}=0$). The matrix elements of the multi-adjacency matrices are given by 
\bea
A^{\vec{m}}_{ij}=\prod_{\alpha=1}^M\left[a_{ij}^{[\alpha]}m^{[\alpha]}+(1-a_{ij}^{[\alpha]})(1-m^{[\alpha]})\right].
\eea
Since any two nodes can be connected only by a single multilink we have 
\bea
\sum_{\vec{m}}A_{ij}^{\vec{m}}=1.
\eea
Having defined the multi-adjacency matrices, it is possible to introduce the definition of the {\em multidegree} $k_i^{\vec{m}}$ as the sum of multilinks $\vec{m}$ incident to the node $i$ \cite{bianconi2013statistical}, i.e.
\bea
k_i^{\vec{m}}=\sum_{j=1}^NA_{ij}^{\vec{m}}.
\eea
Using the approach described in this work we can either define exchangeable sparse multiplex networks in which each layer is independent of the other and  has a  given degree distribution (eventually dependent on the choice of the layer); or we can define exchangeable sparse multiplex networks in which  the multidegree distribution is kept fixed.    

The first case can be modelled by drawing each layer of the multiplex network independently from an exchangeable ensemble of  uncorrelated simple networks. Given the simplicity of the approach here we neglect to treat this case in detail. The latter case can be modelled by an exchangeable multiplex network ensemble in which  each nodes has a series of non trivial multidegrees ${\bf k}^{\vec{m}}_i$ with $\vec{m}\neq\vec{0}$ [e.g. ${\bf k}^{\vec{m}}_i=(k_i^{(1,0)},k_i^{(0,1)},k_i^{(1,1)})$ in the case of $M=2$ layers] with multidegree distribution $\tilde{\pi}({\bf k}^{\vec{m}}_i)$. Moreover we impose that in the network there are exactly $L^{\vec{m}}=\Avg{k^{\vec{m}}}N/2$ multilinks of type $\vec{m}\neq\vec{0}$ and that the multiplex is sparse, i.e. the multidegree $k^{\vec{m}}_i$ has a minimum value greater or equal than $\hat{m}$ and a maximum value smaller or equal than $K^{\vec{m}}$ with $K^{\vec{m}}\ll K_S^{\vec{m}}=\sqrt{\Avg{k^{\vec{m}}}N}$. 
%\end{document}
Therefore the ensemble is defined by  associating to each multiplex network of $M$ layers $\vec{G}=(G_1,G_2,\ldots, G_M)$ the probability 
%\end{document}
%\end{document}
\bea
\mathbb{P}(\vec{G})=P\left(\{k^{\vec{m}}\}\right)e^{-\Sigma\left(\{ k^{\vec{m}}\}\right)}\prod_{\vec{m}\neq \vec{0}}
\left[\delta\left(L^{\vec{m}},\sum_{i<j}A_{ij}^{\vec{m}}\right)\right.\nonumber \\
\left.\times\theta\left(K^{\vec{m}}-\max_{i=1,\ldots, N}k_i^{\vec{m}}\right)\theta\left(\min_{i=1,\ldots, N}k_i^{\vec{m}}-\hat{m}\right)\right],
\eea\\
where here $\{k^{\vec{m}}\}$  indicates the sequence of all the non trivial mutlidegrees $\vec{m}\neq\vec{0}$ of every node $i$ of the multiplex network, and where the entropy $\Sigma(\{k^{\vec{m}}\})$ is given by \cite{bianconi2013statistical}
\bea
\Sigma\left(\{{k^{\vec{m}}}\}\right)=\ln \left(\prod_{\vec{m}\neq\vec{0}}\frac{(2L^{\vec{m}})!!}{\prod_{i=1}^Nk_i^{\vec{m}}!}\right)+o(N).
\eea
Here $P(\{k^{\vec{m}}\})$ is given by the product of the probability that each nodes has multidegrees ${\bf k}_i^{\vec{m}}$ 
\bea
P(\{k^{\vec{m}}\})=\prod_{i=1}^N\tilde{\pi}\left({\bf k}_i^{\vec{m}}\right).
\eea
This exchangeable multiplex network ensemble is Hamiltonian as the probability $\mathbb{P}(\vec{G})$ can be written as 
%\end{document}
\bea
\mathbb{P}(\vec{G})&=&e^{-H(G)}\prod_{\vec{m}\neq \vec{0}}
\left[\delta\left(L^{\vec{m}},\sum_{i<j}A_{ij}^{\vec{m}}\right)\theta\left(K^{\vec{m}}-\max_{i=1,\ldots, N}k_i^{\vec{m}}\right)\right.\nonumber \\
&&\times\left.\theta\left(\min_{i=1,\ldots, N}k_i^{\vec{m}}-\hat{m}\right)\right],
\eea
with Hamiltonian $H(G)$ given by
\bea
H(G)=-\sum_{i=1}^N\ln \left(\tilde{\pi}\left({\bf k}_i^{\vec{m}}\right)\right)-\sum_{i=1}^N\sum_{\vec{m}\neq\vec{0}}k_i^{\vec{m}}!+\sum_{\vec{m}\neq\vec{0}}\ln((2 L^{\vec{m}})!).\nonumber
\eea
Using a statistical mechanics treatment of this ensemble (see Appendix E) it can be shown that the marginal probability $p_{ij}^{\vec{m}}$ of observing a multink $\vec{m}\neq \vec{0}$ between node $i$ and node $j$ is given by 
%\end{document}
\bea
p_{ij}^{\vec{m}}=\Avg{A_{ij}^{\vec{m}}}=\sum_{{k}^{\vec{m}},{k}^{\prime,\vec{m}}}\tilde{\pi}\left({\bf k}^{\vec{m}}\right)\tilde{\pi}_{\vec{m}}\left({\bf k}^{\prime,\vec{m}}\right)p(k^{\vec{m}},k^{\prime,\vec{m}}),
\label{marginalM}
\eea
where the marginal probability $p(k^{\vec{m}},k^{\prime,\vec{m}})$ of observing a multilink $\vec{m}\neq\vec{0}$ between node $i$ of multidegree $k_i^{\vec{m}}=k^{\vec{m}},$ and node $j$ of  multidegree $k_j^{\vec{m}}=k^{\prime,\vec{m}}$ is given by  
\bea
p_{ij|k_i^{\vec{m}}=k^{\vec{m}},k_j^{\vec{m}}=k^{\prime,\vec{m}}}=p(k^{\vec{m}},k^{\prime,\vec{m}})
=\frac{k^{\vec{m}}(k^{\prime,\vec{m}})}{\Avg{k^{\vec{m}}}N}.
\label{marginalCM}
\eea
%\end{document}

\section{Exchangeable ensemble of sparse simplicial complexes}
In recent years there has been a surge of interest in higher-order networks \cite{bianconi2021higher,battiston2021physics,battiston2020networks} including simplicial complexes and hypegraphs. Higher-order networks are able to capture the higher-order interactions present in a variety of complex systems including  brain networks, social networks and protein-interaction networks.
Few works have proposed network ensembles for {\em labelled} simplicial complexes \cite{courtney2016generalized,zuev2015exponential}. In this section we will   propose and study exchangeable ensembles of sparse uncorrelated and correlated simplicial complexes.
\subsection{Uncorrelated exchangeable ensembles of simplicial complexes}
A $d$-dimensional simplex $\alpha$ is a set of $d+1$ nodes $\alpha=[i_0,i_1,\ldots, i_d]$ and indicates the higher-order interaction existing these nodes.  
 A pure $d$ dimensional simplicial complex $\mathcal{K}$ is formed by a set of $d$-dimensional simplices and by all the lower-dimensional simplices formed any proper subsets of the nodes of these $d$-dimensional simplices.

A  pure $d$-dimensional simplicial complex $\mathcal{K}$ has a  structure that is fully  determined by the adjacency tensor ${\bf a}$ of elements $a_{\alpha}=1$ if the $d$-dimensional simplex $\alpha=[i_0,i_1,\ldots, i_d]$ belongs to the simplicial complex, and with $a_{\alpha}=0$ otherwise.
The generalized degree $k_i$ of the  generic node $i$ \cite{bianconi2015complex,courtney2016generalized} indicates the number of $d$-dimensional simplices incident to the node $i$ and it can be expressed in terms of the adjacency tensor as  
\bea
k_i=\sum_{\alpha\supset i}a_{\alpha}=\sum_{i_1<i_2<\ldots< i_d}a_{i_0i_1i_2\ldots i_d}.
\eea
The ensemble of {\em labelled} pure $d$-dimensional simplicial complexes  with given generalized degree sequence ${\bf k}=(k_1,k_2,\ldots, k_N\}$ has been studied in Ref.\cite{courtney2016generalized}.
Here we consider the exchangeable ensemble of uncorrelated $d$-dimensional simplicial complexes.
We indicate with $P({\bf k})$ the probability assigned to observing a generalized degree sequence ${\bf k}$, with 
\bea
P({\bf k})=\prod_{i=1}^N \left[p(k_i)\theta(K_s-k_i)\theta(\hat{m}-k_i)\right].
\eea
Therefore the probability of the generalized degree sequence ${\bf k}$ factorizes in the product of the probability $p(k_i)$ that each node $i$ has a generalized degree $k_i=k$.
Moreover we consider that the simplicial complexes are sparse, i.e. they have a {\em structural cutoff} \cite{courtney2016generalized}
\bea
K_S=\left(\frac{(\avg{k}N)^d}{d!}\right)^{1/(d+1)}.
\eea
This implies that the generalized degree of the nodes $k_i$ have a maximum value $K\ll K_S$. Finally, we assume that each node as a generalized degree equal or greater than $\hat{m}$.
This ensemble is generated by  associating to each simplicial complex $\mathcal{K}$ the probability $\mathbb{P}(\mathcal{K})$ given by  
\bea
\mathbb{P}({\mathcal K})=P\left({\bf k}\right)e^{-\Sigma\left({\bf k}\right)}\delta\left(S,\sum_{\alpha\in \mathcal{K}}a_{\alpha}\right)
%\theta\left(K_S-\max_{i=1,2\ldots, N}k_i\right)\theta\left(\min_{i=1,2\ldots, N}k_i-m\right).
\eea
where $S=\avg{k}N/(d+1)$ indicates the number of simplices in the simplicial complex and where $\Sigma({\bf k})$ is the entropy of the ensemble with generalized degree sequence ${\bf k}$. 
In presence of the structural cutoff,  the entropy $\Sigma({\bf k})$ of $d$-dimensional simplicial complexes with generalized degree sequence  ${\bf k}$ is given by \cite{courtney2016generalized}
\bea
\Sigma({\bf k})=\ln \left([(\avg{k}N)!]^{d/(d+1)} \frac{1}{\prod_{i=1}^Nk_i!}(d!)^{-\avg{k}N/(d+1)}\right)+o(N).\nonumber
\eea
It follows that the  exchangeable ensemble of $d$-dimensional simplicial complexes can be obtained by considering the Hamiltonian simplicial complex ensemble 
\bea
\mathbb{P}({\mathcal K})&=&e^{-H(G)}e^{-\Sigma\left({\bf k}\right)}\delta\left(S,\sum_{\alpha\in \mathcal{K}}a_{\alpha}\right)\theta\left(K_S-\max_{i=1,2\ldots, N}k_i\right)\nonumber \\
&&\times\theta\left(\min_{i=1,2\ldots, N}k_i-\hat{m}\right).
\eea
with Hamiltonian $H(G)$ given by 
\bea
H(G)=-\sum_{i=1}^N\ln p(k_i)+\Sigma({\bf k}).
\eea
This ensemble is exchangeable and the marginal probability for each simplex $\alpha$ is given by  (see Appendix F for the derivation)
\bea
p_{\alpha}&=&\sum_{\{k_0,k_1,\ldots, k_{d}}\left[\prod_{r=0}^{d}p(k_i)\right] p(k_0,k_1,\ldots, k_d)\nonumber \\
&=&d!\frac{\avg{k}}{N^{d}}.
\eea
where the marginal probability $p(k_0,k_1,\ldots, k_d)=p({\alpha=[i_0,i_2,\ldots, i_d]|k_{i_r}=k_r})$ of a simplex $\alpha=[i_0,i_2,\ldots, i_d]$ with the generic node $i_r$ having degree   $k_{i_r}=k_r$ is given by the uncorrelated expression \cite{courtney2016generalized},
\bea
p({\alpha=[i_0,i_2,\ldots, i_d]|k_{i_r}=k_r})&=&p(k_0,k_1,\ldots, k_d)\nonumber \\
&=&d!\frac{\prod_{r=0}^{d}k_r}{(\avg{k}N)^d}.
\eea

\subsection{Correlated exchangeable simplicial complex ensemble}
The final example of exchangeable ensemble is the ensemble of sparse correlated $d$-dimensional simplicial complexes  in which each node has generalized degree $k$ with probability $p(k)$ and each $d$-simplex between $d+1$ nodes of  generalized degrees  ${\bf k}_{\alpha}=(k_0,k_1,\ldots, k_r)$ contributes to the probability of the simplicial complex by a term 
$Q(k_0,k_1,\ldots k_d)=Q({\bf k}_{\alpha})$ where $Q({\bf k}_{\alpha})$
is invariant under any permutation of its arguments. Here we impose that the total number of $d$-simplices is $S=\avg{k}N/(d+1)$ and that the maximum generalized degree of the simplicial complex is below or equal to $K$ ensuring sparsity and the minimum generalized degree of the simplicial complex is greater or equal to $\hat{m}$. 
To this end, we assign to each simplicial complex $\mathcal{K}$ a probability $\mathbb{P}(\mathcal{K})$ given by
\bea
\mathbb{P}(\mathcal{K})&=&\prod_{\alpha\in \mathcal{K}}
[Q({\bf k}_{\alpha})]^{a_{\alpha}}\prod_{i=1}^Np(k_i)e^{-\Sigma({\bf k})}\delta\left(S,\sum_{\alpha\in \mathcal{K}}a_{\alpha}\right)\nonumber \\
&&\times \theta\left(K-\max_{i=1,\ldots, N}k_i\right)\theta\left(\hat{m}-\min_{i=1,\ldots, N} k_i\right),
\eea
where $\Sigma({\bf k})$ is the entropy of the ensemble of correlated networks with degree sequence ${\bf k}$ that  can be expressed as  
\bea
\Sigma({\bf k})=\ln \left( [(\avg{k}N)!]^{d/(d+1)}(d!)^{-\avg{k}N/(d+1)}\prod_{i=1}^N\frac{[\gamma(k_i)]^{k_i}}{k_i!}\right)+o(N)\nonumber
\label{S_SC_corr}
\eea
where $\gamma(k)$ is defined self-consistently by the equation
%\begin{widetext}
\bea
\hspace*{-5mm}\gamma(k)&=&\frac{1}{\avg{k}^d}\sum_{k_1,k_2,\ldots k_d}Q(k,k_1,k_2,\ldots, k_d)\prod_{r=1}^d\left[p(k_r)\frac{k_r}{\gamma(k_r)}\right].
\label{gamma_SC}
\eea
%\end{widetext}
The marginal probability of this ensemble is given by  the exchangeable expression (see Appendix F for the statistical mechanics derivation)
\bea
p_{\alpha}=\sum_{k_0,k_1,\ldots, k_d}\left[\prod_{r=0}^dp(k_r)\right]p(k_0,k_1,\ldots, k_d)
\eea
with  $p(k_0,k_1,\ldots, k_d)$ expressing the marginal probability of a simplex connecting $d+1$ nodes with degrees $(k_0,k_1,\ldots, k_d)$,
\bea
\hspace*{-5mm}p(k_0,k_1,\ldots, k_d)=\frac{d!}{(\avg{k}N)^d}\sum_{k_0,k_1,k_2,\dots, k_d}Q(k_0,k_1,k_2,\ldots, k_d)\prod_{r=0}^d\left[\frac{k_r}{\gamma(k_r)}\right].\nonumber
\eea

\section{Conclusions}
In this work we propose a statistical mechanics framework able to define  sparse exchangeable network ensembles of a given number of nodes $N$. Here by sparse we mean that the networks have a structural cutoff. This hypothesis is necessary for fully treating the model analytically but it can be removed as long as the entropy $\Sigma({\bf k})$ is known and numerically estimated for every possible degree sequence ${\bf k}$ of the network. The network ensemble can be generated by a simple Metropolis-Hastings algorithm. This statistical mechanics approach  is based on enforcing two  global constraints, such as the total number of links and the value of the  exchangeable Hamiltonian of the ensemble. { Although  every link  has} the same marginal probability, the ensemble can generate networks with very heterogeneous degree distribution. This implies that in order to impose an heterogeneous degree distribution we do not need as for the exponential random graphs to impose an extensive number of local constraints but two global constraints are actually sufficient. 
This approach is here shown to be generalizable to networks with degree correlations, to directed and bipartite networks and to generalized network structures such as multilayer networks and  simplicial complexes. This work provides a physical point of view for addressing the challenging problem of  modelling   exchangeable (but not projective) networks ensembles. The model has wide applications as null model of unlabelled networks, 
{ 
 Indeed in applications it is true that exchangeabiity can be achieved by randomization of the network labels that  can be performed by implementing a label reshuffling procedure, however  our theoretical contribution introduces a new exchangeable network model that  is analytically tractable.  
In physics,  one could think of obtaining  the Maxwell distribution of velocity of the particles of a gas by drawing for each particle a velocity from a Gaussian distribution and then reshuffling the particle labels, yet being able to treat the gas without having to perform the node reshuffling procedure numerically has many advantages. Similarly, here we provide a statistical mechanics and analytically treatable  formulation of exchangeable networks that can be potentially  combined to other network analysis tools coming from statistics, network science or machine learning.
}

In conclusions we hope that this work will stimulate further theoretical and applied research at the frontier between physics, mathematics and applications of network science as the formulation of a sparse exchangeable network model that is also projective would have  applications in a number of fields ranging from data analysis, and machine learning to  sampling of networks,  with profound ramifications in mathematics. 

\bibliography{exchangeable_bib}
%\newpage
%\onecolumn
\appendix

\section{EXCHANGEABLE ENSEMBLE OF SPARSE SIMPLE NETWORKS}
\subsection{Treatment of the exchangeable ensemble of uncorrelated networks }
In this section our goal is to solve the partition function $Z(h)$  (that for construction is expected for $h=0$ to take the value $Z(0)=1$) for the exchangeable ensemble of simple networks given by Eq. () using the saddle point equation deriving the expression of the functional order parameter  $c(\omega,k)$.

The us start by recalling the expression given  in the main text 
 for the partition function $Z(h)$ of this network ensemble,
\bea
Z(h)&=&\sum_{G}\mathbb{P}(G)e^{-h\sum_{i<j}a_{ij}}\nonumber \\
&=&\frac{1}{(2L)!!}\sum_{{\bf a}}\sum_{\bf k}^{\prime}\int {{\mathcal D}\omega}\int \frac{d\lambda}{2\pi} e^{G(\lambda,\bm{\omega},{\bf k},h)},
\label{Zs}
\eea
with 
\bea
G(\lambda,\bm{\omega},{\bf k},h)&=&\sum_{i=1}^N [\mi \omega_i k_i+\ln (k_i!p(k_i))]+\mi\lambda L\nonumber \\
&+&\frac{1}{2}\sum_{i,j}\ln (1+e^{-\mi\lambda-\mi\omega_i-\mi\omega_j-h}),
\eea
and with ${\mathcal D}\omega=\prod_{i=1}^N [d\omega_i/(2\pi)]$. In Eq. (\ref{Zs}) and in the following we use the notation $\sum_{\bf k}^{\prime}$ to indicate the sum over all the possible values of the degree of each node $i$ satisfying $\hat{m}\leq k_i\leq K\ll K_s=\sqrt{\avg{k}N}$. Note that by construction we have $Z(h=0)=1$.//
Let us now introduce the functional order parameter \cite{courtney2016generalized,bianconi2008entropies,monasson1997statistical}
\bea
c(\omega,k)=\frac{1}{N}\sum_{i=1}^N\delta(\omega-\omega_i)\delta(k,k_i),
\eea
by enforcing its definition with a series of delta functions. Therefore, by assuming a discretization in $\omega$ in intervals of size $\Delta \omega$ we introduce for any value of $(\omega, k)$ the term
\begin{widetext}
\bea
&&1=\int dc(\omega,k)\delta\left(c(\omega,k)-\frac{1}{N}\sum_{i=1}^N\delta(\omega-\omega_i)\delta(k,k_i)\right)= \int \frac{d\hat{c}(\omega, k)dc(\omega,k) }{2\pi/(N\Delta \omega)}\exp\left[{\mi\Delta\omega \hat{c}(\omega,k)[Nc(\omega, k)-\sum_{i=1}^N\delta(\omega-\omega_i)\delta(k,k_i)]}\right].\nonumber
\label{deltasc}
\eea
\end{widetext}
After performing these operations, by imposing $2L=\avg{k}N$ where $\avg{k}=\sum_kkp(k)$,  the  partition function reads in the limit $\Delta\omega\to 0$,
\bea
Z(h)=\frac{1}{(2L)!!}\sum_{\bf k}^{\prime}\int {\mathcal D}{c}({\omega,k})\int {\mathcal D}\hat{c}({\omega,k})\int \frac{d\lambda}{2\pi} e^{Nf(\lambda,c({\omega},k),\hat{c}(\omega, k),h)}\nonumber
\eea
with 
\begin{widetext}
\bea
&&f(\lambda,c({\omega},k),\hat{c}(\omega, k),h)=\mi\int d\omega \sum_{\hat{m}\leq k\leq K}\hat{c}(\omega,k)c(\omega, k)+\mi\lambda \avg{k}/2+\Psi+\ln \int \frac{d\omega}{2\pi} \sum_{\hat{m}\leq k\leq K} p(k)k!e^{\mi\omega k-\mi\hat{c}(\omega,k)}
\label{fsimple}
\eea
where $\Psi$ is given by 
\bea
\Psi=\frac{N}{2}\int d\omega \int d\omega' \sum_{\hat{m}\leq k\leq K,\hat{m}\leq k'\leq K} c(\omega,k) c(\omega',k')\ln \left(1+e^{-\mi\lambda-\mi\omega-\mi\omega'-h}\right)\nonumber \label{Psi}
\eea
\end{widetext}
and where ${\mathcal D}c(\omega,k)$ is the functional measure ${\mathcal D}c(\omega,k)=\lim_{\Delta \omega\to 0}\prod_{\omega}\prod_{k}^N[dc(\omega,k) \sqrt{N\Delta\omega/(2\pi)}]$ and similarly ${\mathcal D}\hat{c}(\omega,k)=\lim_{\Delta \omega\to 0}\prod_{\omega}\prod_{k}^N[d\hat{c}(\omega,k) \sqrt{N\Delta\omega/(2\pi)}]$.
Performing a Wick rotation in $\lambda$ and assuming $z/N=e^{-\mi\lambda}$ real and much smaller than one, i.e. $z/N\ll 1$ which is allowed in the sparse regime $K\ll K_S$, we can linearize the logarithm and express $\Psi$ as 
\bea
\Psi=\frac{1}{2}z\nu^2e^{-h},
\eea with 
\bea
\nu=\int d\omega\sum_{\hat{m}\leq k\leq K} c(\omega,k) e^{-\mi\omega}.
\eea
The saddle point equations determining the value of the partition function can be obtained by performing the (functional) derivative of $f(\lambda,c({\omega},k),\hat{c}(\omega, k),h)$ with respect to $c(\omega,k)$, $\hat{c}(\omega,k)$ and $\lambda$, obtaining for $h\to 0$,
\bea
-\mi\hat{c}(\omega,k)&=&z\nu e^{-\mi\omega},\nonumber \\
c(\omega,k)&=&\frac{\frac{1}{2\pi}p(k)k!e^{\mi\omega k-\mi\hat{c}(\omega,k)}}{\int \frac{d\omega'}{2\pi} \sum_{\hat{m}\leq k'\leq K} p(k')k'!e^{\mi\omega' k'-\mi\hat{c}(\omega',k')}},\nonumber \\
z\nu^2&=&\avg{k}.
\eea
Let us first calculate the integral 
\bea
\hspace{-8mm}\int \frac{d\omega}{2\pi} \sum_{\hat{m}\leq k\leq K} p(k)k!e^{-\mi\omega k-\mi\hat{c}(\omega,k)}=\int \frac{d\omega}{2\pi} \sum_{\hat{m}\leq k\leq K} k! p(k)e^{\mi\omega k+z\nu e^{-\mi\omega}}\nonumber
\eea
where we have substituted the saddle point expression for $\hat{c}(\omega,k)$. This integral can be also written as 
\bea
\int \frac{d\omega}{2\pi} \sum_{\hat{m}\leq k\leq K} k! p(k)e^{\mi\omega k}\sum_{h=0}^{\infty}\frac{(z\nu)^h}{h!} e^{-\mi\omega h}&=&\sum_{\hat{m}\leq k\leq K}  p(k) (z\nu)^k\nonumber \\&=&\Avg{(z\nu)^k}.\nonumber
\eea
Therefore $c(\omega,k)$ at the saddle point solution can be expressed as 
\bea
\hspace{-5mm}c(\omega, k)=\frac{1}{2\pi}\frac{k!p(k)e^{\mi\omega k+(z\nu)e^{-\mi\omega}}}{\Avg{(z\nu)^k}}\nonumber
\label{Csimple1}
\eea
With this expression, using a similar procedure  we can express $\nu$ as 
\bea
\nu &=&\int d\omega \sum_{\hat{m}\leq k\leq K} c(\omega,k) e^{-\mi\omega}\nonumber \\
%&=&\frac{1}{\Avg{(z\nu)^k}}\sum_{\hat{m}\leq k\leq K}k{p(k)(z\nu)^{k-1}.
\eea
Combing this equation with the third saddle point equation 
\bea
z\nu^2=\avg{k},
\eea
it is immediate to show that $z\nu=1$ is the solution with 
\bea
z=\frac{1}{\avg{k}},\quad \nu=\avg{k}.
\eea
By inserting this expression in Eq. (\ref{Csimple1}) we get Eq. (\ref{c}), i.e.
\bea
c(\omega,k)=\frac{1}{2\pi}k!p(k)e^{\mi\omega k+e^{-\mi\omega}}.
\eea
Calculating the partition function at the saddle point, we get $Z(h\to 0)=1$. 
%\end{document}
\subsection{Calculation of the marginal probability of a link}
For calculating the marginal probability $p_{ij}$ of a link between node $i$ and node $j$ in the exchangeable network ensemble we first note that given that the ensemble has an exchangeable Hamiltonian, the marginal probability of a link should be the same for every link of the network, i.e. $p_{ij}=\tilde{p}$. In order to obtain $\tilde{p}$ we can simply derive the free energy $F=Nf$ with $f$ given by Eq. (\ref{fsimple}) with respect to the auxiliary field $h$ obtaining

\bea
&&\frac{N(N-1)}{2}\tilde{p}=-\left.\frac{\partial (Nf)}{\partial h}\right|_{h= 0}=-\left.\frac{\partial (N\Psi)}{\partial h}\right|_{h= 0}\nonumber \\&&=\frac{N}{2}z\int d\omega \int d\omega' \sum_{\hat{m}\leq k\leq K;\hat{m}\leq k'\leq K} c(\omega,k) c(\omega',k')e^{-\mi\omega-\mi\omega'}\nonumber
\eea
%\end{document}
from which, inserting the saddle point value of $c(\omega,k)$ and $z$ and performing the integrals we get for $N\gg1$,
\bea
p_{ij}=\tilde{p}=\sum_{\hat{m}\leq k\leq K}\sum_{\hat{m}\leq k'\leq K}p(k)p(k')\frac{kk'}{\avg{k}N}=\frac{\avg{k}}{N}.
\eea

%\end{document}
\subsection{Expression of the marginal probability of a link conditioned on the degrees of its two endnodes}
In this paragraph our goal is to derive the expression of the probability $p_{ij|k_i=k,k_j=k'}=p(k,k')$ of a link between node $i$ and node $j$ in the exchangeable network ensemble conditioned on the degree of the two endnodes.
The expression for $p_{ij|k_i=k,k_j=k'}$ can be obtained by showing that the probability  $\hat{\pi}_{ij}$ that node $i$ is connected to node $j$ in any network ensemble enforcing a given degree sequence ${\bf k}$ (the configuration model) is given by 
\bea
\hat{\pi}_{ij}=\sum_{\bf a}a_{ij}\prod_{r=1}^N\delta\left(k_r-\sum_{s=1}^Na_{rs}\right)\delta\left(L-\sum_{r<s}a_{rs}\right)e^{-\Sigma({\bf k})}=\frac{k_ik_r}{\avg{k}N}\nonumber
\label{pi}
\eea
%\end{document}
as long as the maximum degree of the network $K$ is much smaller than the structural cutoff, i.e. $K\ll K_S$.  Since $\hat{\pi}_{ij}$ only depends on the degrees $k_i$ and $k_j$ of its two endnodes, in this ensemble the probability $\hat{\pi}({k,k'})$ of any link between any two nodes of degree $k$ and degree $k'$ , takes the expression, 
\bea
\hat{\pi}_{ij|k_i=k,k_j=k'}=\hat{\pi}(k,k')=\frac{kk'}{\Avg{k}N},
\label{pi2}
\eea
as long as $K\ll K_S$.
The exchangeable network model is essentially an ensemble in which we can get very different degree distributions but each network $G$ with a given distribution ${\bf k}$ is weighted by $P({\bf k})\exp[-\Sigma({\bf k})]$. Therefore the we can express $p_{ij|k_i=k,k_j=k'}$ as
\bea
p_{ij|k_i=k,k_j=k'}&=&p(k,k')=\frac{\sum_{{\bf k}|{k_i=k,k_j=k'}}\prod_{r=1}^N p(k_r)\hat{\pi}({k_i,k_j})}{p(k)p(k')}\nonumber \\
&=&\hat{\pi}(k,k')=\frac{kk'}{\Avg{k}N}.
\eea 
Let us now derive Eq. (\ref{pi2}) for the ensemble in which we fix the degree sequence of the network (for the other examples of ensemble the derivation is similar and we will omit for space constraints).
%\end{document}
To this end we consider the partition function 
\bea
\tilde{Z}({\bf h})&=&\sum_{\bf a}\exp\left[-{\sum_{i<j}h_{k_i,k_j}a_{ij}}\right]\left[\prod_{r=1}^N\delta\left(k_r-\sum_{s=1}^Na_{rs}\right)\right]\times\nonumber \\&&
\times\delta\left(L-\sum_{r<s}a_{rs}\right)e^{-\Sigma({\bf k})}
\label{Zt}
\eea
%\end{document}
where we have introduced some auxiliary fields ${\bf h}=\{h_{k,k'}\}$ where each different auxiliary field $h_{k,k'}$ is associated to the links between nodes of degree $k$ and degree $k'$. Here  the entropy $\Sigma({\bf k})$ of the network with given degree sequence with $k_i\ll K_S$ obeys the Bender-Canfield formula \cite{bender,bianconi2008entropies,anand2009entropy,anand2010gibbs}
\bea
{\Sigma}({\bf k})= \ln \left(\frac{(2L)!!}{\prod_{i=1}^Nk_i!}\right)+o(N)
\label{S2}
\eea
 where in Eq. (\ref{Zt}) and (\ref{S2}) we indicate with  $k_i$ the degree of node $i$ given by 
$k_i=\sum_{j=1}^Na_{ij},$.
%\end{document}
Expressing the Kronecker delta in Eq. (\ref{Zt}) in integral form we get 
 for the partition function $\tilde{Z}({\bf h})$ of this network ensemble,
\bea
\tilde{Z}({\bf h})=\frac{1}{(2L)!!}\sum_{{\bf a}}\int {{\mathcal D}\omega}\int \frac{d\lambda}{2\pi} e^{\tilde{G}(\lambda,\bm{\omega},{\bf k},{\bf h})},
\label{Zs2}
\eea
with 
\bea
\tilde{G}(\lambda,\bm{\omega},{\bf k},{\bf h})&=&\sum_{i=1}^N [\mi\omega_i k_i+\ln (k_i!)]+\mi\lambda L\nonumber \\
&&+\frac{1}{2}\sum_{i,j}\ln \left(1+e^{-\mi\lambda-\mi\omega_i-\mi\omega_j-h_{k_i,k_j}}\right),
\eea
and with ${\mathcal D}\omega=\prod_{i=1}^N [d\omega_i/(2\pi)]$. In Eq. (\ref{Zs2}).
By indicating with $N_k$ the fraction of nodes with degree $k$, 
let us  introduce the functional order parameters \cite{courtney2016generalized,bianconi2008entropies,monasson1997statistical}
\bea
c_k(\omega)=\frac{1}{N_k}\sum_{i=1}^N\delta(\omega-\omega_i)\delta(k,k_i),
\eea
determining the fraction of nodes of degree $k$ that are associated to  $\omega_i=\omega$. 
Therefore, by assuming a discretization in $\omega$ in intervals of size $\Delta \omega$ we introduce for any value of $\omega$ and $k$ the term
\begin{widetext}
\bea
&&1=\int dc_k(\omega)\delta\left(c_k(\omega)-\frac{1}{N_k}\sum_{i=1}^N\delta(\omega-\omega_i)\delta(k,k_i)\right)= \int \frac{d\hat{c}_k(\omega)d c_k(\omega)}{2\pi/(N_k\Delta \omega)}\exp\left[{\mi\Delta\omega \hat{c}_k(\omega)[N_kc_k(\omega)-\sum_{i=1}^N\delta(\omega-\omega_i)\delta(k,k_i)]}\right].\nonumber
\label{deltasc}
\eea
\end{widetext}
After performing these operations, by imposing $2L=\avg{k}N$ where $\avg{k}=\sum_kkp(k)$,  the  partition function reads in the limit $\Delta\omega\to 0$,
\bea
\tilde{Z}({\bf h})=\frac{1}{(2L)!!}\int {\mathcal D}{c}_k({\omega})\int {\mathcal D}\hat{c}_k({\omega})\int \frac{d\lambda}{2\pi} e^{N\tilde{f}(\lambda,c_k({\omega}),\hat{c}_k(\omega),{\bf k},{\bf h})}\nonumber
\eea
with 
\bea
&&\tilde{f}(\lambda,c_k({\omega}),\hat{c}_k(\omega),{\bf k},{\bf h})=\mi\int d\omega \sum_{\hat{m}\leq k\leq K} \tilde{P}(k)\hat{c}_k(\omega)c_k(\omega)\nonumber \\&+&\mi\lambda \avg{k}/2+\Psi+\sum_{\hat{m}\leq k\leq K}\tilde{P}(k)\ln \int \frac{d\omega}{2\pi}  k!e^{\mi\omega k-\mi\hat{c}_k(\omega)},
\eea
where we have indicated with $\tilde{P}(k)=N_k/N$ and where $\Psi$ is given by 
\begin{widetext}
\bea
\Psi=\frac{N}{2}\sum_{\hat{m}\leq k\leq K,\hat{m}\leq k'\leq K} \tilde{P}(k) \tilde{P}(k')\int d\omega \int d\omega' c_k(\omega) c_{k'}(\omega')\ln \left(1+e^{-\mi\lambda-\mi\omega-\mi\omega'-h_{k,k'}}\right)\nonumber 
\eea
\end{widetext}
and where ${\mathcal D}c_k(\omega)$ is the functional measure ${\mathcal D}c_k(\omega)=\lim_{\Delta \omega\to 0}\prod_{\omega}\prod_{k}[dc_k(\omega) \sqrt{N_k\Delta\omega/(2\pi)}]$ and similarly ${\mathcal D}\hat{c}_k(\omega)=\lim_{\Delta \omega\to 0}\prod_{\omega}\prod_{k}[d\hat{c}_k(\omega) \sqrt{N_k\Delta\omega/(2\pi)}]$.
Performing a Wick rotation in $\lambda$ and assuming $z/N=e^{-i\lambda}$ real and much smaller than one, i.e. $z/N\ll 1$ which is allowed in the sparse regime $K\ll K_S$, we can linearize the logarithm and express $\Psi$ as 
\bea
\Psi=\frac{1}{2}z\sum_{\hat{m}\leq k\leq K}\sum_{\hat{m}\leq k'\leq K}\tilde{P}(k) \tilde{P}(k')\nu_k\nu_{k'} e^{-h_{k,k'}},
\eea with 
\bea
\nu_k=\int d\omega c_k(\omega) e^{-\mi\omega}.
\eea
For later convenience let us also define $\nu$ as 
\bea
\nu=\sum_{\hat{m}\leq k\leq K} \tilde{P}(k)\int d\omega c_k(\omega) e^{-\mi\omega}.
\eea
The saddle point equations determining the value of the partition function can be obtained by performing the (functional) derivative of $f(\lambda,c_k({\omega}),\hat{c}_k(\omega),{\bf k},{\bf h})$ with respect to $c_k(\omega)$, $\hat{c}_k(\omega)$ and $\lambda$, obtaining for $h_{k,k'}\to 0$,
\bea
-\mi\hat{c}_k(\omega)&=&z\nu e^{-\mi\omega},\nonumber \\
c_k(\omega)&=&\frac{\frac{1}{2\pi}k!e^{\mi\omega k-\mi\hat{c}_k(\omega)}}{\int \frac{d\omega'}{2\pi}  k!e^{\mi\omega' k-\mi\hat{c}_{k}(\omega')}},\nonumber \\
z\nu^2&=&\avg{k}.
\eea
Let us first calculate the integral 
\bea
\hspace{-8mm}\int \frac{d\omega}{2\pi}  k!e^{-\mi\omega k-\mi\hat{c_k}(\omega)}=\int \frac{d\omega}{2\pi}  k! e^{\mi\omega k+z\nu e^{-\mi\omega}}
\eea
where we have substituted the saddle point expression for $\hat{c}_k(\omega)$. This integral can be also written as 
\bea
\hspace{-8mm}\int \frac{d\omega}{2\pi} e^{\mi\omega k}\sum_{h=0}^{\infty}\frac{(z\nu)^h}{h!} e^{-\mi\omega h}=  (z\nu)^k.
\eea
Therefore $c_k(\omega)$ at the saddle point solution can be expressed as 
\bea
c(\omega, k)=\frac{1}{2\pi}\frac{k!e^{\mi\omega k+(z\nu)e^{-\mi\omega}}}{{(z\nu)^k}}
\label{Csimple2}
\eea
With this expression, using a similar procedure  we can express $\nu$ as 
\bea
\nu&=&\int d\omega \sum_{m\le k\leq K} \tilde{P}(k) {c}_k(\omega)e^{-\mi\omega}\nonumber \\&=&\sum_{\hat{m}\leq k\leq K}\frac{k\tilde{P}(k)}{(z\nu)}=\frac{\avg{k}}{z\nu}
\eea
Therefore this equation reduces to the  third saddle point equation 
\bea
z\nu^2=\avg{k},
\eea
it is immediate to show that $z\nu=1$ is a solution with 
\bea
z=\frac{1}{\avg{k}},\quad \nu=\avg{k}.
\eea
By inserting this expression in Eq. (\ref{Csimple2}) we get Eq. (\ref{c}), i.e.
\bea
c_k(\omega)=\frac{1}{2\pi}k!e^{\mi\omega k+e^{-\mi\omega}}.
\eea
The marginal probability $\hat{\pi}({k,k'})$ of a link between a node of degree $k$ and a node of degree $k'$ can be expressed as 
\bea
N_kN_{k'}\hat{\pi}({k,k'})=\left.\frac{\partial N\tilde{f}}{\partial h_{k,k'}}\right|_{{\bf h}={\bf 0}}
\eea
leading to 
\bea
\hat{\pi}({k,k'})=\frac{z}{N}\int d\omega \int d\omega' c_k(\omega) c_{k'}(\omega')e^{-\mi\omega-\mi\omega'}=\frac{kk'}{\avg{k}N}.\nonumber
\eea
It follows that 
\bea
p(k,k')=\hat{\pi}(k,k')=\frac{kk'}{\avg{k}N}.
\eea
\section{EXCHANGEABLE ENSEMBLE OF SPARSE SIMPLE NETWORKS WITH DEGREE CORRELATIONS}
\subsection{Treatment of the exchangeable ensemble of sparse correlated simple networks}
In this section our goal is to treat the exchangeable ensemble of sparse correlated networks in which each node has degree $k$ with probability $p(k)$ and each link between a node of degree $k$ and a node of degree $k'$ contributes to the partition function by a term  $Q(k,k')=Q(k',k)$. Here we impose that the total number of links $L=\avg{k}N/2$ and that the maximum degree of the network is below or equal to $K$ and the minimum degree of the network is greater or equal to $\hat{m}$. For simplicity of notation we take the auxiliary field $h=0$ from the beginning and we express the   partition function $Z$ of the exchangeable ensemble of sparse correlated networks  as 
\bea
Z=\sum_{{\bf a}}\sum_{{\bf k}}^{\prime}\prod_{i<j}Q(k_i,k_j)]^{a_{ij}}e^{-\Sigma({\bf k})}\prod_{i=1}^N\delta\left(k_i,\sum_{j=1}^Na_{ij}\right)\delta\left(L,\sum_{i<j}a_{ij}\right),\nonumber
\eea
%\end{document}
with the entropy $\Sigma({\bf k})$ given by 
\bea
\Sigma({\bf k})=\ln \left((2L)!! \prod_{i=1}^N\frac{[\gamma(k_i)]^{k_i}}{k_i!}\right)+o(N)
\eea
where $\gamma(k)$ is determined by the self-consistent equation
\bea
\gamma(k)=\frac{1}{\avg{k}}\sum_{\hat{m}\leq k'\leq K}Q(k,k')p(k')\frac{k'}{\gamma(k')}.
\label{gamma_1}
\eea
By expressing the Kronecker deltas in integral form 
\bea
\delta(x,y)=\frac{1}{2\pi}\int_{-\pi}^{\pi}d\omega e^{\mi\omega(x-y)}
\label{delta2}
\eea
we get 
\bea
Z=\frac{1}{(2L)!!}\sum_{\bf k}^{\prime}\int {\mathcal D}\bm\omega \int \frac{d\lambda}{2\pi} e^{G(\bm\omega,\lambda,{\bf k})}
\eea
where $G(\bm\omega,\lambda,{\bf k})$ is given by 
\bea
G(\bm\omega,\lambda,{\bf k})&=&\sum_{i=1}^N[\mi\omega_i k_i+\ln (k_i!p(k_i))-k_i\ln \gamma(k_i)]+\mi\lambda L\nonumber \\&&+\frac{1}{2}\sum_{i,j}\ln \left(1+Q(k_i,k_j)e^{-\mi\lambda-\mi\omega_i-\mi\omega_j}\right),
\eea
and where ${\mathcal D}\omega=\prod_{i=1}^N [d\omega_i/(2\pi)]$.
Let us now introduce the functional order parameter \cite{courtney2016generalized,bianconi2008entropies,monasson1997statistical}
\bea
c(\omega,k)=\frac{1}{N}\sum_{i=1}^N\delta(\omega-\omega_i)\delta(k,k_i),
\eea
by enforcing its definition with a series of delta functions. Therefore, by assuming a discretization in $\omega$ in intervals of size $\Delta \omega$ we introduce for every $(\omega,k)$ the term
\begin{widetext}
\bea
&&1=\int dc(\omega,k)\delta\left(c(\omega,k)-\frac{1}{N}\sum_{i=1}^N\delta(\omega-\omega_i)\delta(k,k_i)\right)=\int \frac{d\hat{c}(\omega, k)dc(\omega,k)}{2\pi/(N\Delta \omega)}e^{\mi\Delta\omega\hat{c}(\omega,k)[Nc(\omega, k)-\sum_{i=1}^N\delta(\omega-\omega_i)\delta(k,k_i)]}
\label{deltasc3}
\eea
\end{widetext}
After performing these operations, by imposing $2L=\avg{k}N$ where $\avg{k}=\sum_kkp(k)$,  the  partition function reads in the limit $\Delta \omega\to 0$
\bea
Z=\frac{1}{(2L)!!}\sum_{\bf k}^{\prime}\int {\mathcal D}{c}({\omega,k})\int {\mathcal D}\hat{c}({\omega,k})\int \frac{d\lambda}{2\pi} e^{Nf(\lambda,c({\omega},k),\hat{c}(\omega, k))}\nonumber
\eea
with 
\bea
&&f(\lambda,c({\omega},k),\hat{c}(\omega, k))=\mi\int d\omega \sum_{\hat{m}\leq k\leq K}\hat{c}(\omega,k)c(\omega, k)\nonumber \\&&+\mi\lambda \avg{k}/2+\Psi+\ln \int d\omega \sum_{\hat{m}\leq k\leq K} p(k)\frac{k!}{[\gamma(k)]^k}e^{\mi\omega k-\mi\hat{c}(\omega,k)}
\eea
where $\Psi$ is given by 
\begin{widetext}
\bea
\Psi&=&\frac{N}{2}\int d\omega \int d\omega' \sum_{\hat{m}\leq k\leq K,\hat{m}\leq k'\leq K}c(\omega,k)c(\omega',k')Q(k,k')\ln \left(1+e^{-\mi\lambda-\mi\omega-\mi\omega'}\right),
\eea
\end{widetext}
 where ${\mathcal D}c(\omega,k)$ and ${\mathcal D}\hat{c}(\omega,k)$ have the same definition then in the simple uncorrelated case.
Performing a Wick rotation in $\lambda$ and assuming $z/N=e^{-\mi\lambda}$ real and much smaller than one, i.e. $z/N\ll 1$ which is allowed in the sparse regime $K\ll K_S$, we can linearize the logarithm and express $\Psi$ as 
\bea
\Psi&=&\frac{z}{2}\int d\omega \int d\omega' \sum_{\hat{m}\leq k\leq K,\hat{m}\leq k'\leq K}c(\omega,k)c(\omega',k')Q(k,k')e^{-\mi\omega-\mi\omega'},\nonumber
\eea
The saddle point equations determining the value of the partition function  read
\bea
&&-\mi\hat{c}(\omega,k)=z e^{-\mi\omega}\int d\omega' \sum_{\hat{m}\leq k'\leq K}Q(k,k')c(\omega',k')e^{-\mi\omega'}\\
&&c(\omega,k)=\frac{\frac{1}{2\pi}p(k)\frac{k!}{[\gamma(k)]^k}e^{\mi\omega k-\mi\hat{c}(\omega,k)}}{\frac{1}{2\pi}\int d\omega' \sum_{\hat{m}\leq k'\leq K} p(k')\frac{k'!}{[\gamma(k')]^{k'}} e^{\mi\omega' k'-\mi\hat{c}(\omega',k')}}\nonumber \\
&&z\int d\omega \int d\omega' \sum_{\hat{m}\leq k\leq K,\hat{m}\leq k'\leq K}c(\omega,k)c(\omega',k')Q(k,k')e^{-\mi\omega-\mi\omega'}=\avg{k}\nonumber
\eea
Let us define $\tilde{\gamma}(k)$ as 
\bea
\tilde{\gamma}(k)=z \int d\omega' \sum_{\hat{m}\leq k'\leq K}Q(k,k')c(\omega',k')e^{-\mi\omega'}.
\eea
With this definition we have 
\bea
-i\hat{c}(\omega,k)=\tilde{\gamma}(k)e^{-\mi\omega}
\eea
Let us first calculate the integral 
\begin{widetext}
\bea
\hspace{-8mm}\frac{1}{2\pi}\int d\omega \sum_{\hat{m}\leq k\leq K} p(k)\frac{k!}{[\gamma(k)]^k}e^{\mi\omega k-\mi\hat{c}(\omega,k)}=\frac{1}{2\pi}\int d\omega \sum_{\hat{m}\leq k\leq K} \frac{k!}{[\gamma(k)]^k} p(k)e^{\mi\omega k+\tilde{\gamma}(k) e^{-\mi\omega}}
\eea
\end{widetext}
where we have substitute the saddle point expression for $\hat{c}(\omega,k)$. This integral can be also written as 
\bea
\hspace{-8mm}\int d\omega \sum_{\hat{m}\leq k \leq K} \frac{k!}{\gamma(k)^k} p(k)e^{\mi\omega k}\sum_{h=0}^{\infty}\frac{(\tilde{\gamma}(k))^h}{h!} e^{-\mi\omega h}=\sum_{\hat{m}\leq k \leq K}  p(k)\left(\frac{\tilde{\gamma}(k)}{\gamma(k)}\right)^k.\nonumber
\eea
Let $w$ indicate the value of this integral , i.e.
\bea
w=\sum_{\hat{m}\leq k\leq K}  p(k)\left(\frac{\tilde{\gamma}(k)}{\gamma(k)}\right)^k.
\eea
The functional order parameter  $c(\omega,k)$ at the saddle point solution can be expressed as 
\bea
c(\omega, k)=\frac{1}{2\pi w}\frac{k!p(k)}{{[\gamma(k)]^k}}e^{\mi\omega k+\tilde{\gamma}(k)e^{-\mi\omega}}
\eea
With this expression, using a similar procedure  we can express $\tilde{\gamma}(k)$ as 
\bea
\tilde{\gamma}(k)&=&z\int d\omega' \sum_{\hat{m}\leq k'\leq K} Q(k,k'){c}(\omega',k')e^{-\mi\omega'}\nonumber \\&=&\frac{z}{w}\sum_{m \leq k'\leq K}Q(k,k')p(k')\frac{k'}{\tilde\gamma(k')}\left(\frac{\tilde\gamma(k')}{\gamma(k')}\right)^{k'}.
\eea
Combing this equation with the third saddle point equation  we get
\bea
&z\int d\omega\int d\omega' \sum_{\hat{m}\leq k\leq K}\sum_{\hat{m}\leq k'\leq K} Q(k,k'){c}(\omega,k){c}(\omega',k')e^{-\mi\omega-\mi\omega'}\nonumber\\ &=\frac{1}{w}
\sum_{m \leq k\leq K} p(k)k\left(\frac{\tilde\gamma(k)}{\gamma(k)}\right)^k=\avg{k},
\eea
Given that $\gamma(k)$ is defined though the Eq. (\ref{gamma_1}),  it follows that 
\bea
\tilde{\gamma}(k)=\gamma(k), \quad w=1,\quad z=\frac{1}{\avg{k}}.
\label{g2}
\eea

Finally using Eqs. (\ref{g2})  we can derive the final expression for  $c(\omega,k)$  given by 
\bea
c(\omega, k)=\frac{1}{2\pi}\frac{k!p(k)}{{[\gamma(k)]^k}}e^{\mi\omega k+{\gamma}(k)e^{-\mi\omega}}.
\eea
From this equation of the functional order parameter we can derive the marginal for each link of the network which is given by 
\bea
p_{ij}=\frac{1}{N}\int d\omega\int d\omega' \sum_{\hat{m}\leq k\leq K,\hat{m}\leq k'\leq K}c(\omega,k)c(\omega',k')Q(k,k')e^{-\mi\omega-\mi\omega'},\nonumber\eea
yielding,
\bea
p_{ij}=\sum_{\hat{m}\leq k\leq K,\hat{m}\leq k'\leq K}p(k)p(k')p(k,k').
\eea
Here $p(k,k')$ indicates the probability of a link between node $i$ and node $j$ conditioned to the degree of the two nodes $k_i=k$ and $k_j=k'$, i.e.
\bea
p_{ij|k_i=k,k_j=k'}=p(k,k')=\frac{1}{\avg{k}N}Q(k,k')\frac{kk'}{\gamma(k)\gamma(k')}.
\eea
Note that for  $Q(k,k')=1$ it follows that $\gamma(k)=1$, and for $Q(k,k')=kk'$ it follows $\gamma(k)=k$ and hence in both cases we recover the exchangeable network ensemble of simple uncorrelated networks.
\\

\section{EXCHANGEABLE ENSEMBLE OF SPARSE DIRECTED NETWORKS}

\subsection{Derivation of the marginal probability}

In this section our goal is the solve the partition function $Z$ for the exchangeable ensemble of directed networks  using the saddle point equation the expression for the marginal probability of a link.
 For simplicity for this ensemble we put the auxiliary fields $h=0$ from the beginning and we express the  partition function $Z$ as 
\bea
Z&=&\sum_{{\bf a}}\sum_{{\bf k}^{in}}'\sum_{{\bf k}^{out}}'e^{-H(G)}\prod_{i=1}^N \left[\delta\left(k_i^{in}-\sum_{j=1}^Na_{ji}\right)\delta\left(k_i^{out}-\sum_{j=1}^Na_{ij}\right)\right]\nonumber \\&&\times \delta\left(L,\sum_{i,j}a_{ij}\right).
\eea
By expressing the Kronecker deltas in integral form 
\bea
\delta(x,y)=\frac{1}{2\pi}\int_{-\pi}^{\pi}d\omega e^{\mi\omega(x-y)}
\label{delta2}
\eea
we get 
\bea
Z=\frac{1}{L!}\sum_{{\bf a}}\sum_{{\bf k}^{in}}^{\prime}\sum_{{\bf k}^{out}}^{\prime}\int {{\mathcal D}\omega}\int {{\mathcal D}\hat{\omega}}\int \frac{d\lambda}{2\pi} e^{G(\lambda,\bm{\omega},\bm{\hat{\omega}},{\bf k}^{in},{\bf k}^{out})},\nonumber 
\eea
with 
\bea
G(\lambda,\bm{\omega},\bm{\hat{\omega}},{\bf k}^{in},{\bf k}^{out})&=&\sum_{i=1}^N [\mi\omega_i k_i^{in}+\mi\hat{\omega}_ik_i^{out}+\ln (k_i^{in}!k_i^{out}!p_d(k_i^{in},k^{out}_i))]\nonumber \\&&+\mi\lambda L
+\sum_{i,j}\ln (1+e^{-\mi\lambda-\mi\omega_i-\mi\hat{\omega}_j}),
\eea
and with ${\mathcal D}\omega=\prod_{i=1}^N [d\omega_i/(2\pi)]$, and ${\mathcal D}\hat\omega=\prod_{i=1}^N [d\hat\omega_i/(2\pi)].$
Let us now introduce the functional order parameter \cite{courtney2016generalized,bianconi2008entropies,monasson1997statistical}
\bea
c(\omega,\hat{\omega},k^{in},k^{out})=\frac{1}{N}\sum_{i=1}^N\delta(\omega-\omega_i)\delta(\hat\omega-\hat\omega_i)\delta(k^{in},k_i^{in})\delta(k^{out},k_i^{out}),\nonumber
\eea
by enforcing its definition with a series of delta functions by introducing the conjugated order parameter $\hat{c}(\omega,\hat{\omega},k^{in},k^{out})$ and by imposing $L=\avg{k^{in}}N=\avg{k^{out}}N$ where $\avg{k^{in}}=\sum_{k^{in},k^{out}}k^{in}p_d(k^{in},k^{out})$, $\avg{k^{out}}=\sum_{k^{in},k^{out}}k^{out}p_d(k^{in},k^{out})$,  the  partition function reads 
\begin{widetext}
\bea
Z=\frac{1}{L!}\sum_{{\bf k}^{in}}^{\prime}\sum_{{\bf k}^{out}}^{\prime}\int {\mathcal D}{c}({\omega,\hat\omega,k^{in},k^{out}})\int {\mathcal D}\hat{c}({\omega,\hat\omega,k^{in},k^{out}})\int \frac{d\lambda}{2\pi} e^{Nf}
\eea
\end{widetext}
with 
\begin{widetext}
\bea
f&=&\mi\int d\omega \int d\hat{\omega}\sum_{\hat{m}\leq k^{in}\leq K,\hat{m}\leq k^{out}\leq K}\hat{c}(\omega,\hat{\omega},k^{in},k^{out})c(\omega,\hat{\omega}, k^{in},k^{out})+\mi\lambda \avg{k^{in}}\nonumber \\
&&+\Psi+\ln \int \frac{d\omega}{2\pi} \sum_{\hat{m}\leq k^{in}\leq K;\hat{m}\leq k^{out}\leq K} p_d(k^{in},k^{out})k^{in}!k^{out}!\exp\left[{\mi\omega k^{in}+\mi\hat{\omega}k^{out}-\mi\hat{c}(\omega,\hat{\omega},k^{in},k^{out})}\right]
\eea
%\end{widetext}
where $\Psi$ is given by 
%\begin{widetext}
\bea
\Psi=\frac{N}{2}\int d\omega \int d\omega' \sum_{\hat{m}\leq k^{in}\leq K,\hat{m}\leq k^{out}\leq K} \sum_{\hat{m}\leq k^{',in}\leq K,\hat{m}\leq k^{',out}\leq K} c(\omega,\hat{\omega},k^{in},k^{out}) c(\omega',\hat{\omega}',k^{',in},k^{',out})\ln \left(1+e^{-\mi\lambda-\mi\omega-\mi\hat{\omega'}}\right)\nonumber 
\eea
\end{widetext}
and where ${\mathcal D}{c}({\omega,\hat\omega,k^{in},k^{out}})$ and ${\mathcal D}\hat{c}({\omega,\hat\omega,k^{in},k^{out}})$ are functional measures.
Performing a Wick rotation in $\lambda$ and assuming $z/N=e^{-\mi\lambda}$ real and much smaller than one, i.e. $z/N\ll 1$ which is allowed in the sparse regime $K\ll K_S$, we can linearize the logarithm and express $\Psi$ as 
%\begin{widetext}
\bea
\Psi=z\nu\hat{\nu},
\eea with 
\bea
\nu=\int d\omega\int d\hat{\omega}\sum_{\hat{m}\leq k^{in}\leq K;\hat{m}\leq k^{out}\leq K} c(\omega,\hat{\omega},k^{in},k^{out}) e^{-\mi\omega},\nonumber \\
\hat{\nu}=\int d\omega\int d\hat{\omega}\sum_{\hat{m}\leq k^{in}\leq K;\hat{m}\leq k^{out}\leq K} c(\omega,\hat{\omega},k^{in},k^{out}) e^{-\mi\hat{\omega}}.\nonumber \\
\eea
%\end{widetext}
The saddle point equations determining the value of the partition function can be obtained by performing the (functional) derivative of $f(\lambda,c({\omega},k),\hat{c}(\omega, k))$ with respect to $c(\omega,\hat{\omega},k^{in},k^{out})$, $\hat{c}(\omega,\hat{\omega},k^{in},k^{ouy})$ and $\lambda$, obtaining
\begin{widetext}
\bea
-\mi\hat{c}(\omega,\hat{\omega},k^{in},k^{out})&=&z\hat{\nu} e^{-\mi\omega}+z\nu e^{-\mi\hat{\omega}},\nonumber \\
c(\omega,\hat{\omega},k^{in},k^{out})&=&\frac{\frac{1}{(2\pi)^2}p_d(k^{in},k^{out})k^{in}!k^{out!}\exp\left[{\mi\omega k^{in}+\mi\hat{\omega}k^{out}-\mi\hat{c}(\omega,\hat{\omega},k^{in},k^{out})}\right]}{\int \frac{d\omega'}{2\pi} \int \frac{d\hat{\omega}'}{2\pi}\sum_{\hat{m}\leq k^{',in}\leq K;\hat{m}\leq k^{',out}\leq K} p_d(k^{',in},k^{',out})k^{',in}!k^{',out}!\exp\left[{\mi\omega' k^{',in}+\mi\hat\omega' k^{',out}-\mi\hat{c}(\omega',\hat\omega',k^{',in},k^{',out})}\right]},\nonumber \\
z\nu{\hat\nu}&=&\avg{k^{in}}.
\eea
%\end{widetext}
Let us first calculate the integral 
%\begin{widetext}
\bea
\int \frac{d\omega}{2\pi} \int \frac{d\hat{\omega}}{2\pi}\sum_{\hat{m}\leq k^{in}\leq K;\hat{m}\leq k^{out}\leq K} p_d(k^{in},k^{out})k^{in}!k^{out}!\exp\left[{\mi\omega k^{in}+\mi\hat\omega k^{out}-\mi\hat{c}(\omega,\hat\omega,k^{in},k^{out})}\right]
\eea
\end{widetext}
where we have substituted the saddle point expression for $\hat{c}(\omega,k)$. Using expanding the exponential and proceeding as in the simple uncorrelated case we get  
\begin{widetext}
\bea
\int \frac{d\omega}{2\pi} \int \frac{d\hat{\omega}}{2\pi}\sum_{\hat{m}\leq k^{in}\leq K;\hat{m}\leq k^{out}\leq K} p_d(k^{in},k^{out})k^{in}!k^{out}!\exp\left[{\mi\omega k^{in}+\mi\hat\omega k^{out}-\mi\hat{c}(\omega,\hat\omega,k^{in},k^{out})}\right]&=&\sum_{\hat{m}\leq k^{in}\leq K} \sum_{\hat{m}\leq k^{out}\leq K}  p_d(k^{in},k^{out})(z\hat{\nu})^{k^{in}}(z\nu)^{k^{out}}\nonumber \\
&=&\Avg{(z\hat{\nu})^{k^{in}}(z\nu)^{k^{out}}}.
\eea
\end{widetext}
Therefore $c(\omega,k)$ at the saddle point solution can be expressed as 
\bea
c(\omega, k)=\frac{1}{(2\pi)^2}\frac{k^{in}!k^{out}!p_d(k^{in},k^{out})e^{\mi\omega k+(z\hat{\nu})e^{-\mi\omega}+(z{\nu})e^{-\mi\hat\omega}}}{\Avg{(z\hat{\nu})^{k^{in}}(z{\nu})^{k^{out}}}}
\label{Cdirected3}
\eea
With this expression, using a similar procedure  we can express $\nu$ as 
\bea
\nu=\frac{1}{\Avg{(z\hat{\nu})^{k^{in}}(z{\nu})^{k^{out}}}}\sum_{\hat{m}\leq k^{in}\leq K}\sum_{\hat{m}\leq k^{out}\leq K}k^{in}p_d(k^{in},k^{out})(z\hat{\nu})^{k^{in}-1}(z{\nu})^{k^{out}}\nonumber \\
\hat{\nu}=\frac{1}{\Avg{(z\hat{\nu})^{k^{in}}(z{\nu})^{k^{out}}}}\sum_{\hat{m}\leq k^{in}\leq K}\sum_{\hat{m}\leq k^{out}\leq K}k^{out}p_d(k^{in},k^{out})(z\hat{\nu})^{k^{in}}(z{\nu})^{k^{out}-1}\nonumber \\
\eea
Combing this equation with the third saddle point equation 
\bea
z\nu\hat{\nu}=\Avg{k^{in}}=\Avg{k^{out}},
\eea
it is immediate to show that $z\nu=z\hat{\nu}=1$ is a solution with 
\bea
z=\frac{1}{\Avg{k^{in}}},\quad \nu=\hat\nu=\Avg{k^{in}}=\Avg{k^{out}}.
\eea
By inserting this expression in Eq. (\ref{Cdirected3}) we get 
\begin{widetext}
\bea
c(\omega,\hat{\omega},k^{in},k^{out})=\frac{1}{(2\pi)^2}k^{in}!k^{out}!p_d(k^{in},k^{out})\exp\left[{\mi\omega k^{in}+\mi\hat{\omega}k^{out}+e^{-i\omega}+e^{-\mi\hat\omega}}\right].
\eea
\end{widetext}
From this equation we can conclude that the networks of these ensemble have heterogeneous degree distribution, as the density of nodes of in-degree $k^{in}$ and out-degree $k^{out}$ is given the desired joint probability distribution, i.e.
\bea
\int d\omega \int d\hat\omega c(\omega,\hat{\omega}, k^{in}, k^{out}) = p_d(k^{in},k^{out}).
\eea
However the marginal for each link is the same and given by Eq. (\ref{marginal_directed}) with the marginal probability of a link conditioned on the degrees of its two endnodes be given by Eq. (\ref{directed_conditioned}).
\section{EXCHANGEABLE ENSEMBLE OF SPARSE BIPARTITE NETWORKS}

\subsection{Derivation of the marginal probability}
In this section our goal is the solve the partition function $Z$ for the exchangeable ensemble of bipartite networks  using the saddle point equation the expression for the marginal probability of a link.
The  partition function $Z$ of this network ensemble,where for simplicity we have put the auxiliary field $h=0$ from the beginning is given by 
\bea
Z&=&\sum_{{\bf a}}\sum_{{\bf k}}'\sum_{{\bf q}}'e^{-H(G)}\delta\left(L,\sum_{i,\mu}a_{i\mu}\right)\prod_{i=1}^N \left[\delta\left(k_i-\sum_{\mu=1}^Mb_{i\mu}\right)\right]\nonumber \\&&\times\prod_{\mu=1}^M \left[\delta\left(q_{\mu}-\sum_{i=1}^Nb_{i\mu}\right).\right]
\eea
By expressing the Kronecker deltas in integral form 
\bea
\delta(x,y)=\frac{1}{2\pi}\int_{-\pi}^{\pi}d\omega e^{\mi\omega(x-y)}
\label{delta5}
\eea
we get 
\bea
Z=\sum_{{\bf a}}\mathbb{P}(G)=\frac{1}{L!}\sum_{{\bf a}}\sum_{\bf k}^{\prime}\sum_{\bf q}^{\prime}\int {{\mathcal D}\omega}\int {{\mathcal D}\hat{\omega}}\int \frac{d\lambda}{2\pi} e^{G(\lambda,\bm{\omega},\bm{\hat{\omega}},{\bf k},{\bf q})},\nonumber
\eea
with 
\bea
G(\lambda,\bm{\omega},\bm{\hat{\omega}},{\bf k},{\bf q})&=&\sum_{i=1}^N [\mi\omega_i k_i+\ln (k_i!p(k_i)]\nonumber \\&&+\sum_{\mu=1}^M[\mi\hat{\omega}_\mu q_\mu+\ln (q_{\mu}!\hat{p}(q_\mu))]+\mi\lambda L\nonumber \\
&&+\sum_{i,\mu}\ln (1+e^{-\mi\lambda-\mi\omega_i-\mi\hat{\omega}_\mu}),
\eea
and with ${\mathcal D}\omega=\prod_{i=1}^N [d\omega_i/(2\pi)]$, and ${\mathcal D}\hat\omega=\prod_{\mu=1}^M [d\hat\omega_i/(2\pi)].$
Let us now introduce the two functional order parameters \cite{courtney2016generalized,bianconi2008entropies,monasson1997statistical}
\bea
c(\omega,k)=\frac{1}{N}\sum_{i=1}^N\delta(\omega-\omega_i)\delta(k,k_i),\nonumber \\
\sigma(\hat\omega,q)=\frac{1}{M}\sum_{\mu=1}^M\delta(\hat\omega-\hat\omega_\mu)\delta(q,q_\mu),
\eea
by enforcing there definition with a series of delta functions involving $c(\omega,k)$ and $\sigma(\omega,k)$ and their conjugated order parameters $\hat{c}(\omega,k)$ and $\hat{\sigma}(\omega,k)$.
%\bea
%&&\delta\left(c(\omega,k)-\frac{1}{N}\sum_{i=1}^N\delta(\omega-\omega_i)\delta(k,k_i)\right)=\int_{-\pi}^{\pi} \frac{d\hat{c}(\omega, k)}{2\pi/N}\exp\left[{i\hat{c}(\omega,k)\left(Nc(\omega,k)-\sum_{i=1}^N\delta(\omega-\omega_i)\delta(k,k_i)\right)}\right], \nonumber \\
%&&\delta\left(\sigma(\hat\omega,q)-\frac{1}{m}\sum_{\mu=1}^M\delta(\hat\omega-\hat\omega_i)\delta(q,q_\mu)\right)=\int_{-\pi}^{\pi} \frac{d\hat{\sigma}(\hat\omega, q)}{2\pi/N}\exp\left[{i\hat{\sigma}(\hat\omega,q)\left(N\sigma(\hat\omega,q)-\sum_{\mu=1}^M\delta(\hat\omega-\hat\omega_\mu)\delta(q,q_\mu)\right)}\right],
%\eea
Imposing also $L=\avg{k}N=\avg{q}M$ where $\avg{k}=\sum_{k}kp(k)$, $\avg{q}=\sum_{q}q\hat{p}(q)$,  the  partition function reads 
\begin{widetext}
\bea
Z=\frac{1}{L!}\sum_{\bf k}^{\prime}\sum_{\bf q}^{\prime}\int {\mathcal D}\hat{c}({\omega,k})\int {\mathcal D}{c}({\omega,k})\int {\mathcal D}\hat{\sigma}({\hat\omega,q})\int {\mathcal D}{\sigma}({\hat\omega,q})\int \frac{d\lambda}{2\pi} e^{Nf}
\eea
\end{widetext}
with 
\begin{widetext}
\bea
f&=&\mi\int d\omega \sum_{\hat{m}\leq k\leq K}\hat{c}(\omega,k)c(\omega,k)+i\alpha \int d\hat\omega \sum_{\hat{m}\leq q\leq \hat{K}}\hat{\sigma}(\hat\omega,q)\sigma(\hat\omega,q)+\mi\lambda \avg{k}\nonumber \\
&&+\Psi+\ln \int \frac{d\omega}{2\pi} \sum_{\tilde{m}\leq k\leq K} p(k)k!\exp\left[{\mi\omega k-\mi\hat{c}(\omega,k)}\right]+\alpha\ln \int \frac{d\hat\omega}{2\pi} \sum_{\hat{m}\leq q\leq \hat{K}} \hat{p}(q)q!\exp\left[{\mi\hat{\omega}q-\mi\hat{\sigma}(\hat\omega,q)}\right]
\eea
\end{widetext}
and $\Psi$ is given by 
\bea
\Psi=\frac{\alpha N}{2}\int d\omega \int d\hat\omega \sum_{\tilde{m}\leq k\leq K,\hat{m}\leq q\leq \hat{K}} c(\omega,k) \sigma(\hat{\omega},q)\ln \left(1+e^{-\mi\lambda-\mi\omega-\mi\hat{\omega}}\right),\nonumber 
\eea
 where ${\mathcal D}\hat{c}({\omega,k})$, ${\mathcal D}{c}({\omega,k})$, ${\mathcal D}\hat{\sigma}({\hat\omega,q})$ and $ {\mathcal D}{\sigma}({\hat\omega,q})$ are functional measures.
Performing a Wick rotation in $\lambda$ and assuming $z/N=e^{-\mi\lambda}$ real and much smaller than one, i.e. $z/N\ll 1$ which is allowed in the sparse regime $K\ll K_S$, we can linearize the logarithm and express $\Psi$ as 
\bea
\Psi=z\alpha\nu\hat{\nu},
\eea with 
\bea
\nu=\int d\omega\sum_{\hat{m}\leq k\leq K} c(\omega,k) e^{-\mi\omega},\nonumber \\
\hat{\nu}=\int d\hat{\omega}\sum_{\hat{m}\leq q\leq \hat{K}} \sigma(\hat\omega,q) e^{-\mi\hat{\omega}}.\nonumber \\
\eea
The saddle point equations determining the value of the partition function can be obtained by performing the (functional) derivative of $f(\lambda,c({\omega},k),\hat{c}(\omega, k))$ with respect to $c(\omega,\hat{\omega},k^{in},k^{out})$, $\hat{c}(\omega,\hat{\omega},k^{in},k^{out})$ and $\lambda$, obtaining
\bea
-\mi\hat{c}(\omega,k)&=&\alpha z\hat{\nu} e^{-\mi\omega},\nonumber \\
c(\omega,k)&=&\frac{\frac{1}{(2\pi)}p(k)k!\exp\left[\mi\omega k-\mi\hat{c}(\omega,k)\right]}{\int \frac{d\omega'}{2\pi}\sum_{m\leq k'\leq K} p(k')k'!\exp\left[{\mi\omega' k'-\mi\hat{c}(\omega',k')}\right]},\nonumber \\
-\mi\hat{\sigma}(\hat\omega,q)&=& z\nu e^{-\mi\hat{\omega}},\nonumber \\
\sigma(\hat\omega,q)&=&\frac{\frac{1}{2\pi}\hat{p}(q)q!\exp\left[\mi\hat\omega q-\mi\hat{\sigma}(\hat\omega,q)\right]}{\int \frac{d\hat\omega'}{2\pi}\sum_{\hat{m}\leq q'\leq \hat{K}} \tilde{p}(q')q'!\exp\left[{\mi\hat\omega' q'-\mi\hat{\sigma}(\hat\omega',q')}\right]},\nonumber \\
z\alpha\nu{\hat\nu}&=&\avg{k}.
\eea
Let us first calculate the integrals 
\bea
\int \frac{d\omega}{2\pi} \sum_{\hat{m}\leq k\leq K} p(k)k!\exp\left[\mi\omega k-\mi\hat{c}(\omega,k)\right]&=&\sum_{\tilde{m}\leq k\leq K}p(k)(\alpha z\hat{\nu})^k\nonumber \\
&=&\Avg{(\alpha z\hat{\nu})^k}\nonumber \\
 \int \frac{d\hat{\omega}}{2\pi}\sum_{\hat{m}\leq q\leq \hat{K}} \hat{p}(q)q!\exp\left[\mi\hat\omega q-\mi\hat{\sigma}(\hat\omega,q)\right]&=&\sum_{\hat{m}\leq q\leq \hat{K}}\hat{p}(q)( z\hat{\nu})^q\nonumber \\&=&\Avg{(z{\nu})^q}
\eea
where we have substituted the saddle point expression for $\hat{c}(\omega,k)$ and $\hat{\sigma}(\hat{\omega},q)$ and we have followed the same procedure as for calculating the correspodning integrals in the previous case. 
It follows that  $c(\omega,k)$  and $\sigma(\hat{\omega},q)$ at the saddle point solution can be expressed as 
\bea
c(\omega, k)=\frac{1}{2\pi}\frac{k!p(k)e^{\mi\omega k+(\alpha z\hat{\nu})e^{-\mi\omega}}}{\Avg{(\alpha z\hat{\nu})^{k}}}\nonumber \\
\sigma(\hat\omega, q)=\frac{\alpha}{2\pi}\frac{q!\hat{p}(q)e^{\mi\hat{\omega} q+(z{\nu})e^{-\mi\hat\omega}}}{\Avg{(z{\nu})^{q}}}\nonumber \\
\label{Cbipartite}
\eea
With this expression, using a similar procedure as in the precedent integrals, we can express $\nu$ as 
\bea
\nu=\frac{1}{\Avg{(\alpha z\hat{\nu})^{k}}}\sum_{\tilde{m}\leq k\leq K}kp(k)(\alpha z\hat{\nu})^{k-1},\nonumber \\
\hat{\nu}=\frac{1}{\Avg{( z{\nu})^{q}}}\sum_{\hat{m}\leq q\leq \hat{K}}q\hat{p}(q)(z{\nu})^{q-1}.
\eea
Combing this equation with the third saddle point equation 
\bea
\alpha z\nu\hat{\nu}=\Avg{k}=\alpha\Avg{q},
\eea
it is immediate to show that  $z\nu=\alpha z\hat{\nu}=1$ is a solution with 
\bea
 z=\frac{1}{\Avg{k}},\quad \nu=\Avg{k}, \quad \hat\nu=\Avg{q}.
\eea
By inserting this expression in Eq. (\ref{Cbipartite}) we get
\bea
c(\omega,k)=\frac{1}{2\pi}k!p(k)\exp\left[\mi\omega k+e^{-\mi\omega}\right],\nonumber \\
\sigma(\hat\omega,q)=\frac{1}{2\pi}q!\hat{p}(q)\exp\left[\mi\hat\omega q+e^{-\mi\hat\omega}\right],\nonumber \\
\eea
From this equation we can conclude that the networks of these ensemble have heterogeneous degree distribution, as the density of nodes in $V$ with degree $k$ is given by $p(k)$ while the density of nodes in $U$ having degree $q$ is given by $\hat{p}(q)$, i.e.
\bea
\int d\omega \int d\hat\omega c(\omega,\hat{\omega},k)=p(k)\nonumber \\.
\int d\omega \int d\hat\omega \sigma(\omega,\hat{\omega},1)=\hat{p}(q).
\eea
However the marginal for each link is the same and given by Eq. (\ref{marginal_bipartite}) with the marginal probability of a link conditioned on the degrees of its two endnodes be given by Eq. (\ref{bipartite_conditioned}).

\section{EXCHANGEABLE ENSEMBLE OF SPARSE MULTIPLEX NETWORKS}

\subsection{Treatment of the exchangeable ensemble of sparse multiplex networks }
In this section our goal is the solve the partition function $Z$  for the exchangeable ensemble of multiplex  networks.
The  partition function $Z$ of this multiplex network ensemble, is given by 
\bea
Z(\bf h)&=&\sum_{\bf A}\sum_{\{{\bf k}^{\vec{m}}\}}'
e^{-H(G)}e^{-\sum_{ij}\sum_{\vec{m}\neq\vec{0}}h^{\vec{m}}A_{ij}^{\vec{m}}}\prod_{\vec{m}\neq \vec{0}}\left[\delta\left(L^{\vec{m}},\sum_{i,j}A_{ij}^{\vec{m}}\right)\right.\nonumber \\&&\left.\times\prod_{i=1}^N \delta\left(k_i^{\vec{m}}-\sum_{j=1}^NA_{ji}^{\vec{m}}\right)\right].
\eea
Here and  in the following we use the notation $\sum_{\bf k}^{\prime}$ to indicate the sum over all the possible values of the degree of each node $i$ satisfying $\hat{m}\leq k_i^{\vec{m}}\leq K^{\vec{m}}\ll K_s^{\vec{m}}=\sqrt{\Avg{k^{\vec{m}}}N}$.
By expressing the Kronecker deltas in integral form 
\bea
\delta(x,y)=\frac{1}{2\pi}\int_{-\pi}^{\pi}d\omega e^{\mi\omega(x-y)}
\label{deltaM}
\eea
we get  for the partition function $Z$ of this network ensemble,
\bea
\hspace*{-4mm}Z({\bf h})=\frac{1}{\prod_{\vec{m}\neq \vec{0}}(2L^{\vec{m}})!!}\sum_{{\bf A}}\sum_{\{{\bf k}^{\vec{m}}\}}'\int {{\mathcal D}\bm\omega}\int \frac{d\lambda}{2\pi} e^{G(\lambda,\bm{\omega},{\bf k}^{\vec{m}},{\bf h})},
\label{ZsM}
\eea
with 
\bea
&&G(\lambda,\bm{\omega},\{{ k}^{\vec{m}}\},{\bf h})=\sum_{i=1}^N\left\{ \sum_{\vec{m}\neq \vec{0}}\left[\mi\omega_i^{\vec{m}} k_i^{\vec{m}}+\ln (k_i^{\vec{m}}!)\right]+\ln\tilde{\pi}({\bf k}_i^{\vec{m}}))\right\}\nonumber \\
&&
+\mi\sum_{\vec{m}\neq \vec{0}}\lambda^{\vec{m}} L^{\vec{m}}\nonumber 
+\frac{1}{2}\sum_{i,j}\ln \left(1+\sum_{\vec{m}\neq\vec{0}}
e^{-\mi\lambda^{\vec{m}}-\mi\omega_i^{\vec{m}}-\mi\omega_j^{\vec{m}}-h^{\vec{m}}}\right),
\eea
and with ${\mathcal D}\bm\omega=\prod_{i=1}^N \prod_{\vec{m}\neq\vec{0}}[d\omega_i^{\vec{m}}/(2\pi)]$. 
Let us now introduce the functional order parameter \cite{courtney2016generalized,bianconi2008entropies,monasson1997statistical}
\bea
c(\bm \omega,{\bf k}^{\vec{m}})=\frac{1}{N}\sum_{i=1}^N\prod_{\vec{m}\neq \vec{0}}\delta(\omega^{\vec{m}}-\omega_i^{\vec{m}})\delta(k^{\vec{m}},k_i^{\vec{m}}),
\eea
by enforcing its definition with a series of delta functions, 
and  by imposing $2L^{\vec{m}}=\Avg{k^{\vec{m}}}N$ where $\Avg{k^{\vec{m}}}=\sum_{k^{\vec{m}}}k^{\vec{m}}p(k^{\vec{m}})$, we get
\begin{widetext}
\bea
Z({\bf h})=\frac{1}{\prod_{\vec{m}\neq\vec{0}}(2L^{\vec{m}})!!} \sum_{\{{\bf k}^{\vec{m}}\}}'\int {\mathcal D}{c}({\bm\omega,{\bf k}^{\vec{m}}})\int {\mathcal D}\hat{c}({\bm\omega,{\bf k}^{\vec{m}}})\int \frac{d\lambda}{2\pi} e^{Nf(\lambda,c({\bm\omega},{\bf k}^{\vec{m}}),\hat{c}(\bm\omega, {\bf k}^{\vec{m}}),{\bf h})}
\eea
\end{widetext}
with 
\bea
&&f(\lambda,c({\bm\omega},{\bf k}^{\vec{m}}),\hat{c}(\bm\omega,{\bf  k}^{\vec{m}}),{\bf h}))=\mi\int d\bm\omega \sum_{ {\bf k}^{\vec{m}}}^{\prime}\hat{c}(\bm\omega,{\bf k}^{\vec{m}})c(\bm\omega,{\bf  k}^{\vec{m}})\nonumber \\
&&+\mi\sum_{\vec{m}\neq \vec{0}}\lambda^{\vec{m}} \Avg{k^{\vec{m}}}/2+\Psi\nonumber \\
&&+\ln \int \frac{d\bm\omega}{(2\pi)^W} \sum_{ {\bf k}^{\vec{m}}}^{\prime} \tilde{\pi}({\bf k}^{\vec{m}})\left[\prod_{{\vec{m}\neq\vec{0}}}k^{\vec{m}}!e^{\mi\omega^{\vec{m}} k^{\vec{m}}}\right]e^{-\mi\hat{c}(\bm\omega,{\bf k}^{\vec{m}})}
\label{f}
\eea
where $W=2^{M}-1$ indicates the number of non-trivial multilinks $\vec{m}\neq\vec{0}$ and $\Psi$ is given by 
\begin{widetext}
\bea
\Psi=\frac{N}{2}\int d\bm\omega \int d\bm\omega' \sum_{{\bf k}^{\vec{m}},{\bf k}^{\prime,\vec{m}}}^{\prime} c(\bm\omega,{\bf k}^{\vec{m}}) c(\bm\omega',{\bf k}^{\prime,\vec{m}})\ln \left(1+\sum_{\vec{m}\neq\vec{0}}e^{-\mi\lambda^{\vec{m}}-\mi\omega^{\vec{m}}-\mi\omega^{'\vec{m}}-h^{\vec{m}}}\right)\nonumber \label{Psi}
\eea
\end{widetext}
and where ${\mathcal D}c(\bm\omega,{\bf k})$ and  ${\mathcal D}\hat{c}(\bm \omega,{\bf k})$ are functional measures.
Performing a Wick rotation in $\lambda$ and assuming $z^{\vec{m}}/N=e^{-i\lambda^{\vec{m}}}$ real and much smaller than one, i.e. $z^{\vec{m}}/N\ll 1$ which is allowed in the sparse regime $K^{\vec{m}}\ll K_S^{\vec{m}}$, we can linearize the logarithm and express $\Psi$ as 
\bea
\Psi=\frac{1}{2}\sum_{\vec{m}\neq \vec{0}}z^{\vec{m}}[\nu(\vec{m})]^2e^{-h^{\vec{m}}},
\eea with 
\bea
\nu(\vec{m})=\int d\bm\omega\sum_{ {\bf k}^{\vec{m}}}^{\prime} c(\bm\omega,{\bf k}^{\vec{m}}) e^{-\mi\omega^{\vec{m}}}.
\eea
The saddle point equations determining the value of the partition function can be obtained by performing the (functional) derivative of $f(\lambda,c({\bm\omega},{\bf k}^{\vec{m}}),\hat{c}(\bm\omega, {\bf k}^{\vec{m}}))$ with respect to $c(\bm\omega,{\bf k}^{\vec{m}})$, $\hat{c}(\bm\omega,{\bf k}^{\vec{m}})$ and $\lambda^{\vec{m}}$, obtaining for $h^{\vec{m}}\to 0$,
\begin{widetext}
\bea
-\mi\hat{c}(\bm\omega,{\bf k}^{\vec{m}})&=&\sum_{\vec{m}\neq\vec{0}} z^{\vec{m}}\nu(\vec{m})\exp\left[{-\mi\omega^{\vec{m}}-h^{\vec{m}}}\right],\nonumber \\
c(\bm\omega,{\bf k}^{\vec{m}})&=&\frac{\frac{1}{(2\pi)^{W}}\tilde{\pi}({\bf k}^{\vec{m}})\prod_{\vec{m}\neq\vec{0}}\left[k^{\vec{m}}!\exp[\mi\omega^{\vec{m}} k^{\vec{m}}]\right]\exp[-\mi\hat{c}(\bm\omega,{\bf k}^{\vec{m}})]}{\int \frac{d\bm\omega'}{(2\pi)^W} \sum_{ {\bf k}'}^{\prime,\vec{m}} \tilde{\pi}({\bf k}^{'\vec{m}})\prod_{\vec{m}\neq\vec{0}}\left[k^{\prime,\vec{m}}!\exp[\mi\omega^{\prime,\vec{m}} k^{\prime\vec{m}}]\right]\exp[-\mi\hat{c}(\bm\omega',{\bf k}^{\prime,\vec{m}})]},\nonumber \\
z^{\vec{m}}[\nu(\vec{m})]^2&=&\Avg{k^{\vec{m}}}.
\eea
\end{widetext}
By proceeding like in the previous examples, we can perform the integral 
\bea
&&\int \frac{d\bm\omega}{(2\pi)^W} \sum_{ {\bf k}}^{\prime} \tilde{\pi}({\bf k}^{\vec{m}})\prod_{\vec{m}\neq\vec{0}}\left[k^{\vec{m}}!\exp[-\mi\omega^{\vec{m}} k^{\vec{m}}]\right]\exp[-\mi\hat{c}(\bm\omega,{\bf k}^{\vec{m}})]=\nonumber \\&&=\sum_{ {\bf k}^{\vec{m}}}^{\prime}  \tilde{\pi}({\bf k}^{\vec{m}}) \prod_{\vec{m}\neq\vec{0}}\left(z^{\vec{m}}\nu(\vec{m})\right)^{k^{\vec{m}}}=w.
\eea
Therefore $c(\bm\omega,{\bf k}^{\vec{m}})$ at the saddle point solution can be expressed as 
\bea
c(\bm\omega, {\bf k}^{\vec{m}})=\frac{1}{(2\pi)^W w}\tilde{\pi}({\bf k}^{\vec{m}})\prod_{\vec{m}\neq\vec{0}}\exp\left[\mi\omega^{\vec{m}} k^{\vec{m}}+z^{\vec{m}}\nu(\vec{m})e^{-\mi\omega^{\vec{m}}}\right]\nonumber
\label{cM0}
\eea
%\end{document}
With this expression, using a similar procedure  we can express $\nu(\vec{m})$ as 
\bea
&&\nu(\vec{m})=\int d\bm\omega \sum_{\bf k}^{\prime} {c}(\bm\omega,{\bf k}^{\vec{m}})e^{-\mi\omega^{\vec{m}}}\nonumber \\
&&=\frac{1}{a}\sum_{{\bf k}}^{\prime}\tilde{\pi}({\bf k}^{\vec{m}})k^{\vec{m}}
[z^{\vec{m}}\nu({\vec{m}})]^{k^{\vec{m}}-1}\prod_{\vec{m'}\neq\vec{m},\vec{0}}[z^{\vec{m'}}\nu({\vec{m'}})]^{k^{\vec{m'}}}
\eea
Combing this equation with the third saddle point equation 
it is immediate to show that $z^{\vec{m}}\nu(\vec{m})=1$ is a solution with 
\bea
z^{\vec{m}}=\frac{1}{\Avg{k^{\vec{m}}}},\quad \nu(\vec{m})=\Avg{k^{\vec{m}}}\quad w=1.
\eea
By inserting this expression in Eq. (\ref{cM0}) we get 
\bea
c(\bm\omega,{\bf k}^{\vec{m}})=\frac{1}{2\pi}\tilde{\pi}({\bf k}^{\vec{m}})\prod_{\vec{m}\neq\vec{0}}\left\{k^{\vec{m}}!\exp\left[\mi\omega^{\vec{m}} k^{\vec{m}}+e^{-\mi\omega^{\vec{m}}}\right]\right\}.
\eea
From this expression, by proceeding like in the simple network case, we can derive that each node of the network have multidegrees ${\bf k}^{\vec{m}}$ with a probability $\tilde{\pi}({\bf k}^{\vec{m}})$ and that the marginal probability of multilinks are given by Eq. (\ref{marginalM}) and Eq. (\ref{marginalCM}).\\
%\end{document}

\section{EXCHANGEABLE ENSEMBLE OF SPARSE SIMPLICIAL COMPLEXES}

\subsection{Derivation of the marginal probability of a simplex in the uncorrelated exchangeable simplicial complex ensembles}
In this section our goal is the solve the partition function $Z(h)$ (that for construction is expected  to take the value $Z(h=0)=1$) for the exchangeable ensemble of uncorrelated simplicial complexes .
The us start by defying the 
 for the partition function $Z(h)$ of this simplicial complex ensemble,as
\bea
Z(h)&=&\sum_{\bf a}\mathbb{P}(\mathcal{K})e^{-h\sum_{\alpha\in \mathcal{K}}a_{\alpha}}\nonumber \\
&=&\hat{C}\sum_{{\bf a}}\sum_{\bf k}^{\prime}\int {{\mathcal D}\omega}\int \frac{d\lambda}{2\pi} e^{G(\lambda,\bm{\omega},{\bf k},h)},
\label{ZsC}
\eea
with $\hat{C}=\left({[d!]^{\avg{k}N/(d+1)}}/{[(\avg{k}N)!]^{d/(d+1)}}\right)$ and
\bea
G(\lambda,\bm{\omega},{\bf k},h)&=&\sum_{i=1}^N [\mi\omega_i k_i+\ln (k_i!p(k_i))]+\mi\lambda \avg{k}/(d+1)\nonumber \\
&&+\sum_{\alpha\in \mathcal{K}}\ln \left(1+e^{-\mi\sum_{r\subset\alpha}\omega_r-\mi\lambda-h}\right),
\eea
and with ${\mathcal D}\omega=\prod_{i=1}^N [d\omega_i/(2\pi)]$. In Eq. (\ref{ZsC}) and in the following we use the notation $\sum_{\bf k}^{\prime}$ to indicate the sum over all the possible values of the generalized degree of each node $i$ satisfying $m\leq k_i\leq K\ll K_S$.
Let us now introduce the functional order parameter \cite{courtney2016generalized,bianconi2008entropies,monasson1997statistical}
\bea
c(\omega,k)=\frac{1}{N}\sum_{i=1}^N\delta(\omega-\omega_i)\delta(k,k_i),
\eea
by enforcing its definition with a series of delta functions. By assuming a discretization in $\omega$ in intervals of size $\Delta \omega$ we then introduce for each choice of $(\omega,k)$ the term 
\begin{widetext}
\bea
&&1=\int dc(\omega,k)\delta\left(c(\omega,k)-\frac{1}{N}\sum_{i=1}^N\delta(\omega-\omega_i)\delta(k,k_i)\right)= \int \frac{d\hat{c}(\omega, k)dc(\omega,k) }{2\pi/(N\Delta \omega)}\exp\left[{\mi\Delta\omega \hat{c}(\omega,k)[Nc(\omega, k)-\sum_{i=1}^N\delta(\omega-\omega_i)\delta(k,k_i)]}\right].\nonumber
\label{deltasc}
\eea
\end{widetext}
After performing these operations, by imposing $(d+1)S=\avg{k}N$ where $\avg{k}=\sum_kkp(k)$,  the  partition function reads in the limit $\Delta\omega\to 0$,
\begin{widetext}
\bea
Z(h)=\frac{[d!]^{\avg{k}N/(d+1)}}{[(\avg{k}N)!]^{d/(d+1)}}\sum_{\bf k}^{\prime}\int {\mathcal D}{c}({\omega,k})\int {\mathcal D}\hat{c}({\omega,k})\int \frac{d\lambda}{2\pi} e^{Nf(\lambda,c({\omega},k),\hat{c}(\omega, k),h)}
\eea
\end{widetext}
with 
\bea
&&f(\lambda,c({\omega},k),\hat{c}(\omega, k),h)=\mi\int d\omega \sum_k\hat{c}(\omega,k)c(\omega, k)\nonumber \\
&&+\mi\lambda \avg{k}/(d+1)+\Psi+\ln \int \frac{d\omega}{2\pi} \sum_{\hat{m}\leq k\leq K} p(k)k!e^{\mi\omega k-\mi\hat{c}(\omega,k)}\nonumber
\label{fSC}
\eea
where $\Psi$  for $K\ll K_S$ can be approximated  by 
\bea
\Psi=\frac{N^d}{{(d+1)}!}e^{-h-\mi\lambda}\prod_{r=0}^d\left[\sum_{m\leq k_r\leq K} \int d\omega_r c(\omega_r,k_r)  e^{-\mi\omega_r}\right]\nonumber \label{PsiSC}
\eea
and where ${\mathcal D}c(\omega,k)$ is the functional measure ${\mathcal D}c(\omega,k)=\lim_{\Delta \omega\to 0}\prod_{\omega}\prod_{k}^N[dc(\omega,k) \sqrt{N\Delta\omega/(2\pi)}]$ and similarly ${\mathcal D}\hat{c}(\omega,k)=\lim_{\Delta \omega\to 0}\prod_{\omega}\prod_{k}^N[d\hat{c}(\omega,k) \sqrt{N\Delta\omega/(2\pi)}]$.
Performing a Wick rotation in $\lambda$ and assuming $z/N^{d}=e^{-\mi\lambda}$ real and much smaller than one, i.e. $z/N^{d}\ll 1$ which is allowed in the sparse regime $K\ll K_S$, we can linearize the logarithm and express $\Psi$ as 
\bea
\Psi=\frac{1}{(d+1)!}z\nu^{d+1}e^{-h},
\eea with 
\bea
\nu=\int d\omega\sum_{\hat{m}\leq k\leq K} c(\omega,k) e^{-\mi\omega}.
\eea
The saddle point equations determining the value of the partition function can be obtained by performing the (functional) derivative of $f(\lambda,c({\omega},k),\hat{c}(\omega, k),h)$ with respect to $c(\omega,k)$, $\hat{c}(\omega,k)$ and $\lambda$, obtaining for $h\to 0$
\bea
-\mi\hat{c}(\omega,k)&=&\frac{z}{d!}\nu^d e^{-\mi\omega},\nonumber \\
c(\omega,k)&=&\frac{\frac{1}{2\pi}p(k)k!e^{\mi\omega k-\mi\hat{c}(\omega,k)}}{\int \frac{d\omega'}{2\pi} \sum_{m\leq k'\leq K} p(k')k'!e^{\mi\omega' k'-\mi\hat{c}(\omega',k')}},\nonumber \\
\frac{z}{d!}\nu^{d+1}&=&\avg{k}.
\eea
Let us first calculate the integral 
\bea
&&\int \frac{d\omega}{2\pi} \sum_{\hat{m}\leq k\leq K} p(k)k!e^{-\mi\omega k-\mi\hat{c}(\omega,k)}=\nonumber \\
&&=\int \frac{d\omega}{2\pi} \sum_{\hat{m}\leq k\leq K} k! p(k)e^{\mi\omega k+(z\nu^d/d!) e^{-\mi\omega}}
\eea
where we have substituted the saddle point expression for $\hat{c}(\omega,k)$. This integral can be also written as 
\bea
&&\int \frac{d\omega}{2\pi} \sum_{\hat{m}\leq k\leq K} k! p(k)e^{\mi\omega k}\sum_{h=0}^{\infty}\frac{(z\nu^d/d!)^h}{h!} e^{-\mi\omega h}\nonumber \\&&=\sum_{\hat{m}\leq k\leq K}  p(k) \left(\frac{z}{d!}\nu^d\right)^k=\Avg{\left(\frac{z}{d!}\nu^d\right)^k}.
\eea
Therefore $c(\omega,k)$ at the saddle point solution can be expressed as 
\bea
c(\omega, k)=\frac{1}{2\pi}\frac{k!p(k)\exp\left[{\mi\omega k+(z\nu^d/d!)e^{-\mi\omega}}\right]}{\Avg{(z\nu^d/d!)^k}}.
\label{CsimpleS}
\eea
With this expression, using a similar procedure  we can express $\nu$ as 
\bea
\nu=\int d\omega \sum_{k\leq K} {c}(\omega,k)}e^{-\mi\omega}=\frac{1}{\Avg{(z\nu^d/d!)^k}}\sum_{k\leq K}k{p(k)(z\nu^d/d!)^{k-1}.\nonumber
\eea
Combing this equation with the third saddle point equation 
\bea
\frac{z}{d!}\nu^{d+1}=\avg{k},
\eea
it is immediate to show that $z\nu^d/d!=1$ is a solution with 
\bea
z=\frac{d!}{\avg{k}^d},\quad \nu=\avg{k}.
\eea
By inserting this expression in Eq. (\ref{CsimpleS}) we get 
\bea
c(\omega,k)=\frac{1}{2\pi}k!p(k)e^{i\omega k+e^{-i\omega}}.
\eea
Calculating the partition function at the saddle point, we get $Z(h\to 0)=1$. 
%\subsection{Calculation of the marginal probability of a link}
For calculating the marginal distribution $p_{\alpha}$ of a simplex $\alpha$  in the exchangeable network ensemble we first note that given that the ensemble has an exchangeable Hamiltonian, the marginal probability of a simplex should be the same for every simplex of the simplicial complex, i.e. $p_{\alpha}=\tilde{p}$. In order to obtain $\tilde{p}$ we can simply derive the free energy $F=Nf$ with $f$ given by Eq. (\ref{fSC}) with respect to the auxiliary field $h$ obtaining
\bea
\left(\begin{array}{c} N\\d+1\end{array}\right)\tilde{p}&=&-\left.\frac{\partial (Nf)}{\partial h}\right|_{h= 0}=-\left.\frac{\partial (N\Psi)}{\partial h}\right|_{h= 0}\nonumber \\
&=&\frac{N}{(d+1)!}z\left[\int d\omega  \sum_{\hat{m}\leq k\leq K} c(\omega,k)e^{-\mi\omega}\right]^{d+1}
\eea
%\end{document}
from which, by approximating the binomial 
\bea
\left(\begin{array}{c} N\\d+1\end{array}\right)\simeq \frac{N^{d+1}}{(d+1)!}
\eea
for $N\gg1$ and $d$ finite, and
inserting the saddle point value of $c(\omega,k)$ and $z$ we get for $N\gg1$,
\bea
p_{\alpha}=\tilde{p}=\sum_{\{k_r\}|m\leq k_r\leq K}\left[\prod_{r=0}^dp(k_r)\right]p(k_0k_1,\ldots, k_d).
\eea
with 
\bea
p_{\alpha=[i_0,i_1,\ldots, i_d]|k_{i_r}=k_r}=p(k_0k_1,\ldots, k_d)=d!\frac{\prod_{r=0}^dk_r}{(\avg{k}N)^d}.
\eea
\subsection{Derivation of the marginal probability of a simplex in the correlated sparse exchangeable ensemble of simplicial complexes}

The partition function of the exchangeable ensemble of sparse correlated simplicial complexes  can be written as 
\bea
Z({ h})&=&\sum_{{\bf a}}e^{-h\sum_{\alpha}a_{\alpha}}\sum_{{\bf k}}^{\prime}\prod_{\alpha\in \mathcal{K}}Q(k_{i_0},k_{i_1}\ldots, k_{i_d})]^{a_{\alpha}}e^{-\Sigma({\bf k})}\nonumber \\
&&\delta\left(k_i,\sum_{i_1<i_2<\ldots i_d}^Na_{i,i_1,\ldots,i_d}\right)\delta\left(S,\sum_{\alpha\in \mathcal{K}}a_{\alpha}\right),
\eea
with the entropy $\Sigma({\bf k})$ given by 
\bea
\Sigma({\bf k})=\ln \left((\avg{k}N)!]^{d/(d+1)} \frac{1}{(d!)^{-\avg{k}N/(d+1)}}\right)\nonumber \\+
\ln \left( \prod_{i=1}^N\frac{[\gamma(k_i)]^{k_i}}{k_i!}\right)+o(N)
\eea
where $\gamma(k)$ is determined by the self-consistent equation
Eq. (\ref{gamma_simplex})
By expressing the Kronecker deltas in integral form 
\bea
\delta(x,y)=\frac{1}{2\pi}\int_{-\pi}^{\pi}d\omega e^{\mi\omega(x-y)}
\label{delta2}
\eea

we get 
\bea
Z({ h})=\frac{{(d!)^{-\avg{k}N/(d+1)}}}{(\avg{k}N)!]^{d/(d+1)}} \sum_{\bf k}^{\prime}\int {\mathcal D}\bm\omega \int \frac{d\lambda}{2\pi} e^{G(\bm\omega,\lambda,{\bf k},h)}
\eea
%\end{document}
where $G(\bm\omega,\lambda,{\bf k})$ is given by 
\bea
&&G(\bm\omega,\lambda,{\bf k},h)=\sum_{i=1}^N[\mi\omega_i k_i+\ln (k_i!p(k_i))-k_i\ln \gamma(k_i)]\nonumber \\
&&+\mi\lambda \avg{k}/(d+1)+\sum_{\alpha\in \mathcal{K}}\ln \left(1+Q(k_0,k_1,\ldots, k_d)e^{-\mi\lambda-\mi\sum_{r\subset \alpha}\omega_r-h}\right),\nonumber
\eea
and where ${\mathcal D}\omega=\prod_{i=1}^N [d\omega_i/(2\pi)]$.
Let us now introduce the functional order parameter \cite{courtney2016generalized,bianconi2008entropies,monasson1997statistical}
\bea
c(\omega,k)=\frac{1}{N}\sum_{i=1}^N\delta(\omega-\omega_i)\delta(k,k_i),
\eea
by enforcing its definition with a series of delta functions introducing for each choice of $(\omega,k)$ the term 
\begin{widetext}
\bea
&&1=\int dc(\omega,k)\delta\left(c(\omega,k)-\frac{1}{N}\sum_{i=1}^N\delta(\omega-\omega_i)\delta(k,k_i)\right)=\int \frac{d\hat{c}(\omega, k)dc(\omega,k)}{2\pi/(N\Delta \omega)}e^{\mi\Delta \omega\hat{c}(\omega,k)[Nc(\omega, k)-\sum_{i=1}^N\delta(\omega-\omega_i)\delta(k,k_i)]}.
\label{deltasc3}
\eea
\end{widetext}
%\end{document}
After performing these operations, by imposing $(d+1)S=\avg{k}N$ where $\avg{k}=\sum_kkp(k)$,  the  partition function reads 
\bea
Z=\frac{1}{\mathcal{N}}\sum_{\bf k}^{\prime}\int {\mathcal D}{c}({\omega,k})\int {\mathcal D}\hat{c}({\omega,k})\int \frac{d\lambda}{2\pi} e^{Nf(\lambda,c({\omega},k),\hat{c}(\omega, k),h)}\nonumber
\eea
with 
\bea
&&f(\lambda,c({\omega},k),\hat{c}(\omega, k),h)=\mi\int d\omega \sum_k\hat{c}(\omega,k)c(\omega, k)\nonumber \\&&+\mi\lambda \avg{k}/(d+1)+\Psi+\ln \int d\omega \sum_{\hat{m}\leq k\leq K} p(k)\frac{k!}{[\gamma(k)]^k}e^{\mi\omega k-\mi\hat{c}(\omega,k)}\nonumber
\eea
where $\Psi$ is given by 
\begin{widetext}
\bea
\Psi&=&\frac{N^d}{(d+1)!}\sum_{{\bf k}=(k_0,k_1,\ldots,k_d)|m\leq k_r\leq K}\int d\hat{\mathcal D}\omega\prod_rc(\omega_r,k_r)Q(k_0,k_1,\ldots k_d) \ln \left(1+e^{-\mi\lambda-h-\mi\sum_{r\subset \alpha}\omega_r}\right),
\eea
\end{widetext}
 where ${\mathcal D}c(\omega,k)$ and ${\mathcal D}\hat{c}(\omega,k)$ are functional measures.
 %\end{document}
Performing a Wick rotation in $\lambda$ and assuming $z/N^d=e^{-\mi\lambda}$ real and much smaller than one, i.e. $z/N^d\ll 1$ which is allowed in the sparse regime $K\ll K_S$, we can linearize the logarithm and express $\Psi$ as 
\bea
\Psi&=&\frac{z}{(d+1)!}\sum_{{\bf k}=(k_0,k_1,\ldots,k_d)|m\leq k_r\leq K}Q(k_0,k_1,\ldots k_d) \prod_{r=0}^d \nu({k_r}),\nonumber
\eea
where 
\bea
\nu(k)=\int d\omega c(\omega,k)e^{-\mi\omega}.
\eea
The saddle point equations determining the value of the partition function  read for $h\to 0$
 \bea
&&-\mi\hat{c}(\omega,k)= e^{-\mi\omega}\tilde{\gamma}(k)\nonumber\\
&&c(\omega,k)=\frac{\frac{1}{2\pi}p(k)\frac{k!}{[\gamma(k)]^k}e^{\mi\omega k-\mi\hat{c}(\omega,k)}}{\frac{1}{2\pi}\int d\omega' \sum_{\hat{m}\leq k\leq K} p(k)\frac{k!}{[\gamma(k)]^k} e^{\mi\omega' k-\mi\hat{c}(\omega',k)}}\nonumber \\
&&\Psi=\frac{\avg{k}}{d+1}
\eea
where $\tilde{\gamma}(k)$ as 
\bea
\tilde{\gamma}(k)&=&\frac{z}{d!}  \sum_{{\bf k}=(k_1,\ldots, k_d)|\hat{m}\leq k_r\leq K}Q(k,k_1,\ldots,k_d) \nonumber \\
&&\times\prod_{r=1}^d\left[\int d\omega_r c(\omega_r,k_r)e^{-\mi\omega_r}\right].
\eea

Let us first calculate the integral 
\bea
&&\frac{1}{2\pi}\int d\omega \sum_{\hat{m}\leq k\leq K} p(k)\frac{k!}{[\gamma(k)]^k}e^{\mi\omega k-\mi\hat{c}(\omega,k)}\nonumber \\
&&=\frac{1}{2\pi}\int d\omega \sum_{\hat{m}\leq k\leq K} \frac{k!}{[\gamma(k)]^k} p(k)e^{\mi\omega k+\tilde{\gamma}(k) e^{-\mi\omega}}\nonumber
\eea
where we have substitute the saddle point expression for $\hat{c}(\omega,k)$. This integral can be also written as 
\bea
\int d\omega \sum_{\hat{m}\leq k \leq K} \frac{k!}{\gamma(k)^k} p(k)e^{\mi\omega k}\sum_{h=0}^{\infty}\frac{(\tilde{\gamma}(k))^h}{h!} e^{-\mi\omega h}=\sum_{\hat{m}\leq k \leq K}  p(k)\left(\frac{\tilde{\gamma}(k)}{\gamma(k)}\right)^k.\nonumber
\eea
Let $w$ indicate the value of this integral , i.e.
\bea
w=\sum_{\hat{m}\leq k\leq K}  p(k)\left(\frac{\tilde{\gamma}(k)}{\gamma(k)}\right)^k.
\eea
The functional order parameter  $c(\omega,k)$ at the saddle point solution can be expressed as 
\bea
c(\omega, k)=\frac{1}{2\pi w}\frac{k!p(k)}{{[\gamma(k)]^k}}e^{\mi\omega k+\tilde{\gamma}(k)e^{-\mi\omega}}
\eea
With this expression, using a similar procedure  we can express $\tilde{\gamma}(k)$ as 
\bea
\tilde{\gamma}(k)&=&\frac{z}{d!w^d}\sum_{{\bf k}=(k_1,k_2,\ldots, k_r)|\hat{m}\leq k'\leq K} Q(k_0,k_1,\ldots, k_d)\nonumber \\
&&\times\prod_{r=1}^d\left[p(k_r)\frac{k_r}{\tilde\gamma(k_r)}\left(\frac{\tilde\gamma(k_r)}{\gamma(k_r)}\right)^{k_r}\right].
\eea
Combing this equation with the third saddle point equation  we get
\bea
\Psi=\frac{1}{w}
\sum_{m \leq  k\leq K} \left[p(k)k\left(\frac{\tilde\gamma(k)}{\gamma(k)}\right)^k\right]=\avg{k},
\eea
Given that $\gamma(k)$ is defined though the Eq. (\ref{gamma_SC}),  it follows that 
\bea
\tilde{\gamma}(k)=\gamma(k), \quad w=1,\quad z=\frac{d!}{\avg{k}^d}.
\label{g2}
\eea

Finally using Eqs. (\ref{g2})  we can derive the final expression for  $c(\omega,k)$  given by 
\bea
c(\omega, k)=\frac{1}{2\pi}\frac{k!p(k)}{{[\gamma(k)]^k}}e^{\mi\omega k+{\gamma}(k)e^{-\mi\omega}}.
\eea
From this equation of the functional order parameter we can derive the marginal for each link of the network which is given by 
\begin{widetext}
\bea
p_{\alpha}=\frac{d!}{(\avg{k}N)^d}\sum_{{\bf k}=(k_0,k_1,\ldots,k_d)|\hat{m}\leq k_r\leq K}Q(k_0,k_1,\ldots, k_r)\prod_{r=0}^d\left[\int d\omega_r c(\omega_r,k_r)e^{-i\omega_r}\right],\eea
\end{widetext}
yielding,
\bea
p_{\alpha}=\sum_{{\bf k}|\hat{m}\leq k\leq K}\left[\prod_{r=0}^dp(k_r)\right]p(k_0,k_1,\ldots,k_r).
\eea
Here $P(k_0,k_1,\ldots,k_r)$ indicates the probability of a link between node $i$ and node $j$ conditioned to the degree of the two nodes $k_i=k$ and $k_j=k'$, i.e.
\begin{widetext}
\bea
p_{\alpha=[i_0,i_1,\ldots, i_d]|k(i_r)=k_r}=p(k_0,k_1,\ldots, k_r)=\frac{d!}{(\avg{k}N)^d}Q(k_0,k_1,\ldots, k_r)\prod_{r=0}^d\left[\frac{k_r}{\gamma(k_r)}\right].
\eea
\end{widetext}
Note that for  $Q(k_0,k_1,\ldots, k_r)=1$ it follows that $\gamma(k)=1$, and for $Q(k_0,k_1,\ldots,k_r)=\prod_{r=0}^d k_r$ it follows $\gamma(k)=k$ and hence in both cases we recover the exchangeable ensembles of uncorrelated simplicial complexes.
\end{document}